\documentclass[]{jfm}

\usepackage{graphicx}
\usepackage{amsmath}
\usepackage{pgfplots}
\usepackage[normalem]{ulem}
\usepackage{siunitx}
\usepackage{epstopdf}
\usepackage{natbib}
\usepackage{hyperref}
\hypersetup{
	colorlinks = true,
	urlcolor   = blue,
	citecolor  = blue,
}

\newcommand{\RomanNumeralCaps}[1]
\linenumbers

\shorttitle{Transport of small heavy particles in homogeneous turbulence}
\shortauthor{T. Berk and F. Coletti}

\title{Dynamics of small heavy particles in homogeneous turbulence: a Lagrangian experimental study}

\author{Tim Berk\aff{1,}\aff{2}, Filippo Coletti\aff{1,}\aff{2}\corresp{\email{fcoletti@ethz.ch}}\corresp{Present address: Department of Mechanical and Process Engineering, ETH Zurich, Switzerland.}
		}
	
\affiliation{\aff{1}St. Anthony Falls Laboratory, University of Minnesota, Minneapolis, MN 55414, USA
		\aff{2}Department of Aerospace Engineering and Mechanics, University of Minnesota, Minneapolis, MN 55455, USA
	}

\begin{document}

\maketitle   

\begin{abstract}	
We investigate the behavior of microscopic heavy particles settling in homogeneous air turbulence.
The regimes are relevant to the airborne transport of dust and droplets: the Taylor-microscale Reynolds number is $Re_\lambda = 289$ -- 462, the Kolmogorov-scale Stokes number is $St = 1.2$ -- 13, and the Kolmogorov acceleration is comparable to the gravitational acceleration (i.e., the Froude number $Fr = \textit{O}(1)$).
We use high-speed laser imaging to track the particles and simultaneously characterize the air velocity field, resolving all relevant spatio-temporal scales.
The role of the flow sampled by the particles is spotlighted.
In the present range of parameters, the particle settling velocity is enhanced proportionally to the velocity scale of the turbulence.
Both gravity and inertia reduce the velocity fluctuations of the particles compared to the fluid; while they have competing effect on the particle acceleration, through the crossing trajectories and inertial filtering mechanisms, respectively.
The preferential sampling of high-strain/low-vorticity regions is measurable, but its impact on the global statistics is moderate.
The inertial particles have large relative velocity at small separations, which increases their pair dispersion; however, gravity offsets this effect by causing them to experience fluid velocities that decorrelate faster in time compared to tracers.
Based on the observations, we derive an analytical model to predict the particle velocity and acceleration variances for arbitrary $St$, $Fr$, and $Re_\lambda$.
This agrees well with the present observations and previous simulations and captures the respective effects of inertia and gravity, both of which play crucial roles in the transport.
\end{abstract}


\section{Introduction}\label{s:intro}
\subsection{Objective}
The objective of the present study is to explore the effects of inertia and gravity on the motion of heavy particles in homogeneous turbulence.
Given its relevance to countless aspects of natural, industrial, and medical settings, the topic has been studied in depth and a large body of literature is summarized by excellent reviews \citep{Balachandar2010,Gustavsson2016}.
Reaching a predictive understanding of the transport of small airborne particles is even more crucial in the current pandemic \citep{Mittal2020}.
Still, solid evidence concerning important aspects of the particle transport (e.g., fall speed, velocity fluctuations, acceleration, and dispersion) has remained elusive.
The reasons include the difficulty of carrying out detailed measurements resolving all important spatio-temporal scales, the stringent hypotheses of theoretical and numerical models, and the scarcity of one-to-one comparisons between computations and experiments.
In particular, it is noteworthy that the majority of numerical studies have focused on the zero-gravity case.
If this allows to isolate the effect of inertia, it also impedes the direct validation in the laboratory.
Indeed, several behaviors of heavy particles in turbulence were theorized and simulated for decades, e.g., the oversampling of regions of high strain and downward velocity fluctuations \citep{Maxey1987,Squires1991a,Wang1993}; nevertheless, they were only recently demonstrated and quantified by experiments \citep{Petersen2019}.

Here we consider the case of turbulent air laden with solid particles much smaller than the Kolmogorov scale $\eta$.
We focus on the range of Stokes number (based on the Kolmogorov time scale) $St \approx 1$ -- 13, while the acceleration scale of the turbulence is of the same order as the gravitational acceleration.
These conditions are especially relevant to the transport of dust and droplets in the atmosphere.
We perform time-resolved imaging of both the fluid and the particle motion, resolving virtually all scales at play.
The measurements of particle velocity, acceleration, and relative motion highlight the influence of the turbulence and the competing effects of inertia and gravity, demonstrating how in the present regime both effects are crucial for the fate and transport of the dispersed phase.
The analysis culminates with an analytical model that builds on the classic framework put forward by \citet{Csanady1963} to estimate the motion of the particles from the properties of the fluid they sample.
The model is shown to agree well with the present observations, as well as with data in the literature.
The rest of the paper is organized as follows: in \S\ref{ss:background} we summarize the theoretical background and key existing results relevant to the present study; in \S\ref{s:methodology} we describe the experimental apparatus, and the measurement and processing procedure; in \S\ref{s:results} we present and discuss the data; in \S\ref{s:model} we introduce the analytical model and compare it with the observations, before drawing conclusions in \S\ref{s:conclusion}.

\subsection{Background}\label{ss:background}
We briefly review some fundamental relations and concepts that will help interpret and model the behavior of the dispersed phase.
We consider particles much denser than the fluid and sufficiently small compared to any flow scale.
We indicate properties of the particles, fluid, and fluid at the particle location with subscripts ``p'', ``f''and ``fp'', respectively.
If the particle Reynolds number $Re_p$ (based on the particle diameter $d_p$ and a slip velocity from the fluid $u_s$) is sufficiently small, drag and gravity are the only forces usually retained in the equation of motion \citep{Maxey1983}:
\begin{equation}
\boldsymbol{a_p} = \frac{\boldsymbol{u_{fp}} - \boldsymbol{u_p}}{\tau_p} - g \boldsymbol{\hat{e}_y}.
\label{eqn:motion}
\end{equation}
Here $u_p$ is the particle velocity, $a_p = \mathrm{d} u_p/ \mathrm{d} t$ is the particle acceleration, $u_{fp}$ is the fluid velocity at the particle location (which we will refer to as the sampled-fluid velocity), $\tau_p$ is the particle response time, and $g$ is the gravitational acceleration along the unit vector $\boldsymbol{\hat{e}_y}$.
We normalize using the Kolmogorov scales for time, velocity, and acceleration: $\tau_\eta = (\nu/\varepsilon)^{1/2}$, $u_\eta = (\nu \varepsilon)^{1/4}$, and $a_\eta = u_\eta/\tau_\eta =(\varepsilon^3/\nu)^{1/4}$, respectively, where $\nu$ is the kinematic viscosity and $\varepsilon$ is the dissipation rate.
This yields:
\begin{equation}
	\frac{\boldsymbol{a_p}}{a_\eta} = {St}^{-1} \left(\frac{\boldsymbol{u_{fp}}}{u_\eta} - \frac{\boldsymbol{u_{p}}}{u_\eta}\right) - {Fr}^{-1} \boldsymbol{\hat{e}_y}.
	\label{eqn:motion_normalized}
\end{equation}

The Stokes number $St = \tau_p/\tau_\eta$ and the Froude number $Fr = a_\eta/g$ can be combined in the settling parameter $Sv = St/Fr = (\tau_p g)/u_\eta$.
We define the slip velocity as $\boldsymbol{u_s} = \boldsymbol{u_p} - \boldsymbol{u_{fp}}$, i.e. the particle velocity relative to the surrounding flow.
(The opposite sign convention is also used in the literature, defining the slip velocity as the fluid velocity seen by the particle.)
Denoting averaged quantities with angled brackets and fluctuating ones with a prime, the Reynolds decomposition reads:
\begin{equation}
	\boldsymbol{a_p} = \langle \boldsymbol{a_p} \rangle + \boldsymbol{a_p}' = {\tau_p}^{-1} \left(\langle\boldsymbol{u_{fp}}\rangle+\boldsymbol{u_{fp}}'-\langle\boldsymbol{u_{p}}\rangle+\boldsymbol{u_{p}}'\right) - g \boldsymbol{\hat{e}_y}.
\end{equation}
Averaging yields:
\begin{equation}
	\langle\boldsymbol{u_{p}}\rangle - \langle\boldsymbol{u_{fp}}\rangle = \langle\boldsymbol{u_{s}}\rangle = -\tau_p \langle \boldsymbol{a_p} \rangle - \tau_p g \boldsymbol{\hat{e}_y}
	\label{eqn:umean}
\end{equation}
and
\begin{equation}
	\boldsymbol{a_p}' = {\tau_p}^{-1} \left(\boldsymbol{u_{fp}}' - \boldsymbol{u_{p}}'  \right) = -{\tau_p}^{-1} \boldsymbol{u_{s}}'.
	\label{eqn:uprime}
\end{equation}

In equilibrium conditions ($\langle \boldsymbol{a_p} \rangle=0$) and still fluid ($\langle\boldsymbol{u_{fp}}\rangle=0$) the particles settle at a terminal velocity $\langle u_{p,y} \rangle = \langle u_{s,y} \rangle = -\tau_p g$.
Turbulence can either increase or decrease the fall speed through different mechanisms \citep{Nielsen1993,Good2012}.
Perhaps the best-known among those is preferential sweeping, by which inertial particles oversample downward sides of turbulent eddies, leading to a net increase in settling velocity, especially up to $St = \textit{O}(1)$ \citep{Maxey1987,Wang1993}.
Consistently with this view, \citet{Good2014} probed the parameter space and found that settling enhancement was maximum for $St = \textit{O}(1)$ and $Sv = \textit{O}(1)$.
From (\ref{eqn:umean}), the modification of the mean vertical velocity of the particles can be expressed as $\Delta u_y = \langle u_{p,y} \rangle + \tau_p g = \langle u_{p,y} \rangle - \langle u_{s,y} \rangle = \langle u_{fp,y} \rangle$ (negative for settling enhancement).
That is, the settling rate modification is determined by the vertical fluid velocity sampled by the particles.
It is debated which velocity scale governs the phenomenon, with various studies indicating settling enhancement proportional with the root-mean-square (rms) of the fluid velocity fluctuations $u_\mathrm{rms} \equiv \langle (u_f')^2 \rangle^{1/2}$: typically $\Delta u_y \approx - 0.2 u_\mathrm{rms}$ \citep{Aliseda2002,Yang2005,Huck2018}.
Recently \citet{Momenifar2020} showed by theoretical arguments and numerical simulations that the enhancement is in fact governed by a range of velocity scales, whose width increases with $St$.

Averaging over the square of (\ref{eqn:uprime}) leads to:
\begin{equation}
	\langle(\boldsymbol{a_p}')^2\rangle = {\tau_p}^{-2} \left(\langle(\boldsymbol{u_{fp}}')^2\rangle + \langle(\boldsymbol{u_{p}}')^2\rangle - 2 \langle \boldsymbol{u_{fp}}' \boldsymbol{u_{p}}' \rangle \right) = {\tau_p}^{-2} \langle (\boldsymbol{u_{s}}')^2\rangle,	
\end{equation}
which relates the particle velocity and acceleration variance to the sampled-fluid velocity and slip velocity variance.
Of particular relevance to such a relation is the framework developed by \citet{Csanady1963}, who proposed a link between the Lagrangian spectrum of the particle velocity ($E_p$) and the one of the sampled-fluid velocity ($E_{fp}$) through a response function $H^2$:
\begin{equation}
	E_p (\omega) = H^2(\omega) E_{fp}(\omega)
\end{equation}
where $\omega$ is the angular frequency in the Lagrangian frame of reference.
The velocity and acceleration variances can be obtained from the spectra as:
\begin{equation}
	\langle (u_p')^2\rangle = \frac{2}{\upi} \int_0^\infty E_p(\omega)\mathrm{d}\omega
	\label{eqn:uspectra}
\end{equation}
\begin{equation}
	\langle (a_p')^2\rangle = \frac{2}{\upi} \int_0^\infty \omega^2 E_p(\omega)\mathrm{d}\omega
	\label{eqn:aspectra}
\end{equation}
The response function can be modeled from (\ref{eqn:motion}), by taking the Fourier transform of the particle velocity and sampled-fluid velocity \citep{Csanady1963}:
\begin{equation}
	H^2(\omega) = \frac{1}{1+(\omega \tau_p)^2}
	\label{eqn:response}
\end{equation}
\citet{Zhang2019} derived a more complete form of the response function including unsteady forces, which however are expected to be small for microscopic heavy particles.
The dependence of the response function on the particle response time illustrates the particle inability to respond to fluctuations with frequencies greater than $\textit{O}(\tau_p^{-1})$.
This behavior, often termed inertial filtering, was clearly demonstrated in situations where gravity is absent or negligible \citep{Bec2006,Ayyalasomayajula2006}.
In presence of gravity, already \citet{Yudine1959} realized the importance of the drift through turbulent eddies, with particles experiencing fast-changing flow conditions: this crossing-trajectories effect leads to faster decorrelation of the particle motion \citep{Squires1991b,Elghobashi1992,Wang1993a}.
Consequently, gravitational drift enhances particle acceleration compared to zero-gravity conditions, counteracting the effect of inertial filtering \citep{Ireland2016a,Ireland2016b}.
For weakly inertial particles ($St < 1$), preferential sampling of high-strain, low-vorticity regions may also contribute to increasing particle acceleration, as the fluid acceleration is higher in the strain-dominated regions \citep{Bec2006}.

\section{Methodology} \label{s:methodology}
\subsection{Experimental apparatus}
Experiments are performed in a chamber where a region of homogeneous air turbulence is formed by two facing jet arrays.
The facility was introduced and qualified in detail in \citet{Carter2016} and \citet{Carter2017,Carter2018}; here we only give a brief description for completeness.
The chamber measures $2.4 \times 2 \times 1.1$~m$^3$ in $x$, $y$ and $z$ directions (where $x$ is aligned with the jet axis and $y$ is in vertical direction) and has acrylic walls for optical access.
Each jet array consists of 128 quasi-synthetic jets, individually operated according to a sequence proposed by \citet{Variano2008}.
The region of homogeneous turbulence (with negligible mean flow and shear) measures approximately $0.5 \times 0.7 \times 0.4$~m$^3$.
The Reynolds number can be tuned by adjusting the average firing time of the jets ($\mu_{on}$).
In the present study we operate the jets in two modes, and the main turbulence properties for each are reported in table~\ref{tab:1}.
For both cases, the region of homogeneity is substantially larger than the integral length scale of the turbulence, allowing for a natural inter-scale energy cascade without major influence of the boundary conditions.

\begin{table}
	\setlength{\tabcolsep}{6pt}
	\centering
	\begin{tabular}{c c c c c c c c c}
		$\mu_{on}$ (s) & $u_\mathrm{rms}$ (m/s) & $u_{\mathrm{rms},x}/u_{\mathrm{rms},y}$ & $L$ (m) & $T_E$ (s) & $T_L$ (s) & $\eta$ (mm) & $\tau_\eta$ (ms) & $Re_\lambda$ \vspace{.5em}\\
		0.1 & 0.41 & 1.24 & 0.10 & 0.22 & 0.23 & 0.31 & 6.3 & 289\\
		3.2 & 0.73 & 1.86 & 0.14 & 0.23 & 0.23 & 0.24 & 3.6 & 462\\
	\end{tabular}
	\caption{Flow properties for the cases used in the present study. The rms velocity $u_\mathrm{rms}$, integral length scale $L$ and the Eulerian and Lagrangian integral timescales $T_E$ and $T_L$ are based on weighted geometric averages over $x$ and $y$.}
	\label{tab:1}
\end{table}

The materials and procedure followed to investigate the inertial particle transport are similar as in \citet{Petersen2019}.
The particles are fed into the chamber via a 3~m vertical chute connected to the top of the chamber, being released at a steady rate using an AccuRate dry material feeder.
We use three sizes of soda-lime glass beads (density $\rho_p = 2500$~kg/m$^3$), with mean diameters $d_p = 32 \: \pm \: 7 \: \si{\micro\meter}$, $52 \: \pm \: 6 \: \si{\micro\meter}$ and $96 \: \pm \: 11 \: \si{\micro\meter}$, respectively.
Particle response times are evaluated using the Schiller \& Naumann correction \citep{Clift2005}:
\begin{equation}
	\tau_p = \frac{\rho_p d_p^2}{18 \mu (1+0.15 Re_{p,0}^{0.687})}
	\label{eqn:tau_p}
\end{equation}
where $\mu$ is the air dynamic viscosity and $Re_{p,0} = d_p\tau_pg/\nu$ is the particle Reynolds number based on the still-air settling velocity $\tau_p g$.
Iterative evaluation of (\ref{eqn:tau_p}) yields response times of $\tau_p = 7.4$, 17 and 47~ms respectively.
The particles are expected to approach terminal velocity over a distance on the order of $\tau_p^2 g$.
For the largest particles this is 2~cm, an order of magnitude smaller than the extent of the homogeneous region traversed before reaching the measurement field of view.
Table~\ref{tab:2} reports the main non-dimensional parameters for the five experimental cases obtained combining the different particle types and turbulence forcing.
We focus on $St = \textit{O}(1)$ and $Sv = \textit{O}(1)$, with one case of larger $St$ and $Sv$.
For the particle volume fraction and mass fraction in this study (of order $10^{-4}$ and $10^{-7}$, respectively), the flow properties are not expected to be modified by the loading \citep{Petersen2019}.

\subsection{Measurement approach}
Imaging is performed in the $x$--$y$ symmetry plane at the center of the region of homogeneous turbulence.
The flow is seeded with 1--2 \si{\micro\meter} DEHS (di-ethyl-hexyl-sebacate) droplets which faithfully follow the flow.
The flow is illuminated using a Nd:YLF single-pulse laser (Photonics, 30 mJ/pulse) synchronized with a VEO640 camera mounting a 200~mm Nikon lens.
An aperture number $f^\# = 4$ gives a thickness of the focal plane of 1.5~mm.
The active portion of the camera sensor, and therefore the size of the field of view (FOV), depends on the acquisition frequency.
The latter is optimized to give a time separation between consecutive images of at most $0.1 \tau_\eta$, yielding the resolutions reported in table~\ref{tab:3}.
For both $Re_\lambda$, the FOV is much smaller than the region of homogeneous turbulence.
For each case we record 10 separate runs, with total duration of the recordings of around 40 integral time scales.

Data is processed using the procedure detailed in \citet{Petersen2019}.
Raw images are separated in particle-only images and tracer-only images, distinguishing DEHS droplets and glass beads based on brightness and size.
Particle image velocimetry (PIV) is performed on the tracer-only images.
We use an initial interrogation window of 64 by 64 pixels, refined to 32 by 32 pixels with 50~\% overlap, for a vector spacing of 0.9~mm ($2.9\eta$ and $3.7\eta$ for the $Re_\lambda=289$ and 462 cases respectively) that resolves the fine scales of turbulence \citep{Worth2010}.
Particle tracking velocimetry (PTV) is performed on the particle-only images, following the cross-correlation approach.
The fluid velocity is evaluated at the particle locations using weighted linear interpolation of the four neighboring velocity vectors.
A comparison with cubic and spline interpolation shows no significant difference, but linear interpolation substantially decreases the computation time.
Particle velocities and accelerations are determined from the particle position using convolution with the first and second derivative of a Gaussian kernel, respectively.
The width of the latter is chosen as $0.5 \tau_\eta$ and $0.4 \tau_\eta$ for the $Re_\lambda=289$ and 462 cases, respectively, following the procedure established for tracers in \citet{Voth2002} and \citet{Mordant2004} and applied to inertial particles in \citet{Gerashchenko2008}, \citet{Nemes2017} and \citet{Ebrahimian2019a,Ebrahimian2019b}.

The number of samples used to calculate particle statistics ranges between $0.8 \cdot 10^{6}$ and $5.6 \cdot 10^{6}$ for the different cases.
Uncertainty in the statistics is affected by both random uncertainty, due to the finite sample size, and bias uncertainty, due to systematic errors in estimating the particle centroid and the local fluid velocity.
\citet{Baker2021}, who followed a similar time-resolved PIV/PTV approach for particle-laden turbulent boundary layers, showed that the bias uncertainty was negligible compared to the random uncertainty.
Compared to their study, the present particles are much smaller with respect to the flow scales, and therefore the error associated with interpolating the fluid velocity at the particle location is correspondingly smaller.
The centroid location uncertainty (investigated for these same particles in \citet{Petersen2019}) is an order of magnitude smaller than the typical particle displacement as in \citet{Baker2021}, and therefore its impact is also deemed negligible compared to the random uncertainty.
The latter is estimated based on the standard deviation of the last 20~\% of data \citep{Ebrahimian2019b}, and is reported in table~\ref{tab:4} for the main quantities.

\begin{table}
	\setlength{\tabcolsep}{6pt}
	\centering
	\begin{tabular}{c c c c c c c}
		$d_p$ (\si{\micro\meter}) & $Re_\lambda$ & $St$ & $Sv$ & $Fr$ & $d_p/\eta$ & $Re_{p,0}$ \vspace{.5em}\\
		32 & 289 & 1.2 & 1.5 & 0.8 & 0.10 & 0.15\\
		32 & 462 & 2.1 & 1.1 & 1.9 & 0.13 & 0.15\\
		52 & 289 & 2.7 & 3.4 & 0.8 & 0.16 & 0.56\\
		52 & 462 & 4.7 & 2.5 & 1.9 & 0.21 & 0.56\\
		96 & 462 & 13.1 & 6.9 & 1.9 & 0.42 & 3.26\\
	\end{tabular}
	\caption{Relevant non-dimensional parameters for the experimental cases in the present study.}
	\label{tab:2}
\end{table}

\begin{table}
\setlength{\tabcolsep}{6pt}
\centering
\begin{tabular}{c c c c c c}
	$Re_\lambda$ & $f_{aq}$ (Hz) & $\tau_\eta f_{aq}$ & $T_{aq}/T_L$ & FOV ($x \times y$) (cm$^2$) & Resolution (pix/mm) \vspace{.5em}\\
	289 & 1800 & 11.3 & 38 & $11.5 \times 9.0$ & 17.8\\
	462 & 5100 & 18.4 & 44 & $5.8 \times 5.4$ & 17.8\\
\end{tabular}
\caption{Acquisition parameters for the two turbulence forcing cases. $f_{aq}$ and $T_{aq}$ represent the acquisition frequency and length of acquisition, respectively.}
\label{tab:3}
\end{table}

\begin{table}
\setlength{\tabcolsep}{6pt}
\centering
\begin{tabular}{c c c c c c c c}
	$\langle u_{fp,y} \rangle$ & $\langle u_{p,y} \rangle$ & $\langle u_{s,y} \rangle$ & $\langle (u_{fp,y}')^2 \rangle$ & $\langle (u_{p,y}')^2 \rangle$ & $\langle (u_{s,y}')^2 \rangle$ & $\langle (\mathrm{d}u_{fp,y}'/\mathrm{d}t)^2\rangle$ & $\langle (a_{p,y}')^2 \rangle$ \vspace{.5em}\\
	1.4 \% & 0.8 \% & 0.3 \% & 1.3 \% & 1.2 \% & 0.3 \% & 0.3 \% & 0.3 \%\\
\end{tabular}
\caption{Random errors of the velocity and acceleration statistics based on the standard deviation of the last 20 \% of data for the case with the lowest number of samples.}
\label{tab:4}
\end{table}

\section{Results} \label{s:results}
\subsection{Velocity}
\begin{figure}
	\centering
	\includegraphics[]{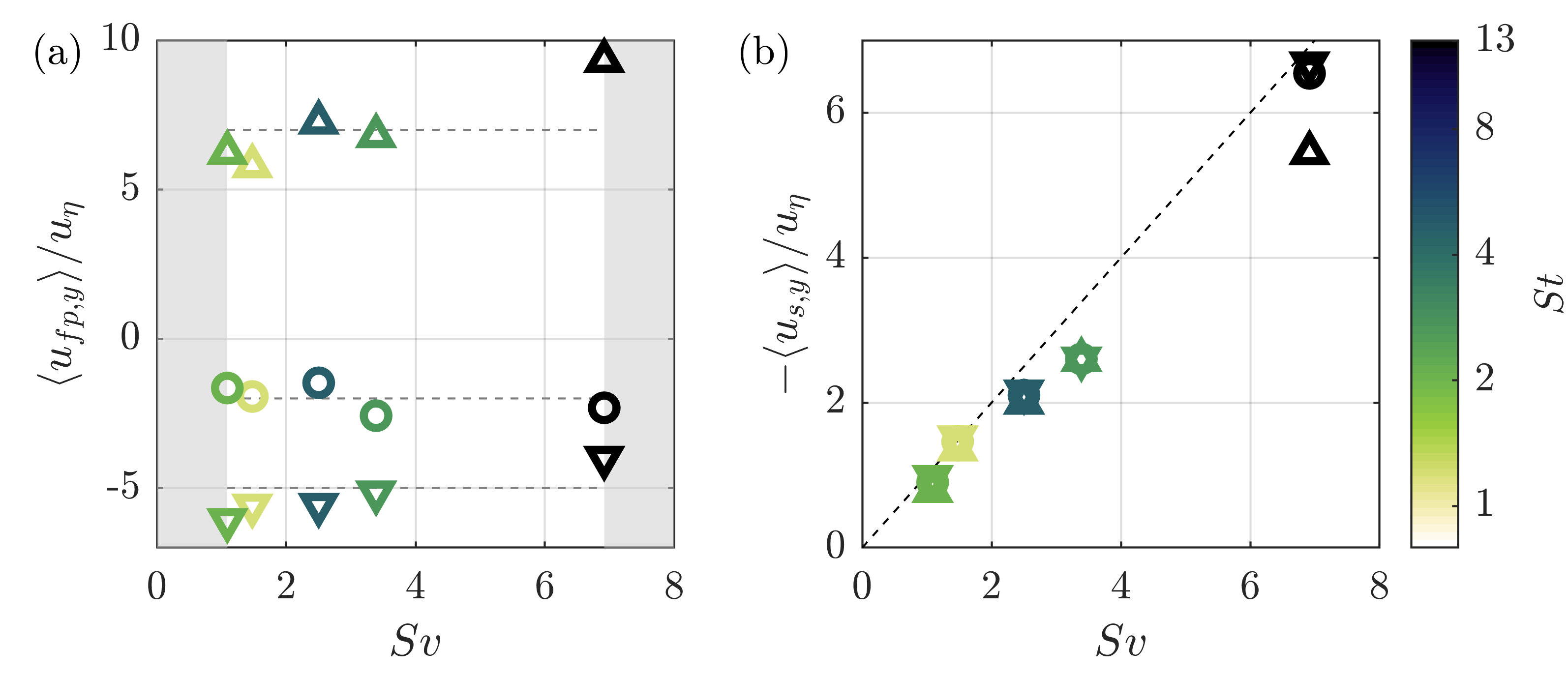}
	\caption{Mean vertical velocity of the sampled flow (a) and mean vertical slip velocity (b). Symbols represent the mean over the upward- and downward-moving subsets (upward- and downward-pointing triangles respectively) as well as the ensemble mean (circles). Dashed lines in (a) represent $\langle u_{fp,y}\rangle/u_\eta = C$, with $C$ defined in the text. The dashed line in (b) represents $\langle u_{s,y} \rangle=-\tau_p g$.
	}
	\label{fig:1}
\end{figure}

We begin by considering the statistics of the sampled fluid, focusing on the vertical component $u_{fp,y}$ which is most relevant to the settling process.
Figure~\ref{fig:1}a shows the mean vertical sampled-fluid velocity for all particles, as well as the velocity conditioned on upward-/downward-moving particles (i.e., particles with positive/negative $u_{fp,y}$).
The results are approximately independent of $Sv$ and consistent with the scaling $\langle u_{fp,y}\rangle/u_\eta = C$, where the constant $C$ is the mean for the respective sets of particles.
Given the expected dependence of the vertical velocities with $St$ and $Sv$, here and in the following figure we shade the area outside of the investigated range to emphasize that the trends should not be extrapolated.
According to previous studies mentioned in \S\ref{s:intro}, we expect $\langle u_{fp,y}\rangle \approx -0.2 u_\mathrm{rms}$, which over the present range of $Re_\lambda$ implies $\langle u_{fp,y}\rangle \approx -2 u_\eta$.
Indeed, for the unconditional average over all particles we find $C \approx -2$.
The results for the downward-moving and upward-moving subsets are consistent with $C \approx -5$ and $C\approx 7$, respectively.
Thus, the upward-/downward-moving particles sample fluid regions with relatively large upward/downward velocity fluctuations, and these are of similar magnitude for both subsets.
Therefore, the net settling enhancement stems from the downward particles being more numerous, not from their association to stronger downward events.

Figure~\ref{fig:1}b shows the normalized slip velocity, the dashed line indicating $\langle u_{s,y} \rangle/u_\eta = -Sv$, or $\langle u_{s,y} \rangle=-\tau_p g$, as theorized by \citet{Wang1993}.
This corresponds well to the measured data, except for the upward moving particles of largest $Sv$ which have a significantly smaller slip velocity.
This is likely related to the assumption $\langle \boldsymbol{a_p} \rangle =0$.
The latter is strictly valid only for the ensemble of all particles, and not necessarily for specific subsets.
In absence of this assumption, (\ref{eqn:umean}) predicts a smaller slip velocity for mean downward particle acceleration.

\begin{figure}
	\centering
	\includegraphics{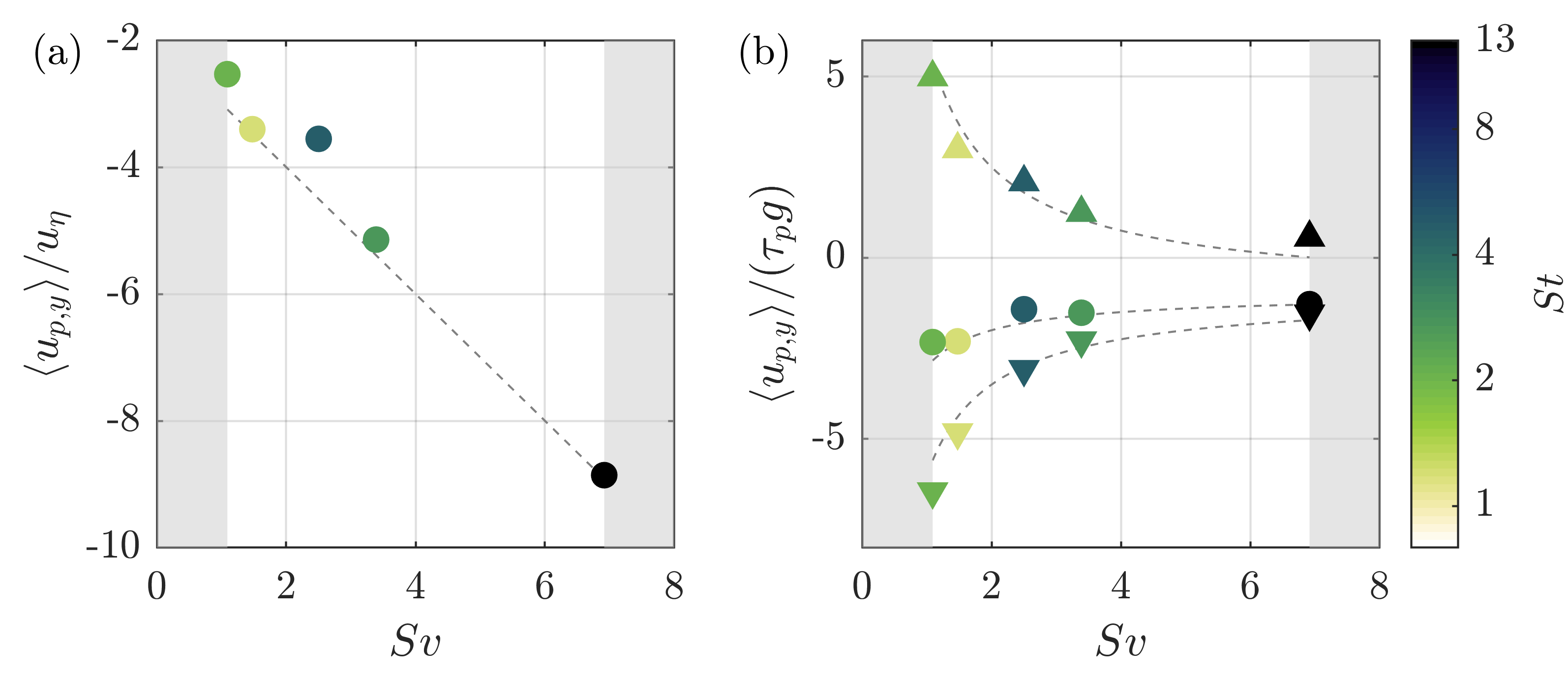}
	\caption{Mean vertical particle velocity, normalized by the Kolmogorov velocity (a) and by the still-air settling velocity $\tau_p g$ (b). Symbols represent the mean over the upward- and downward-moving subsets (upward- and downward-pointing triangles respectively) as well as the ensemble mean (circles).
	The dashed line in (a) represents $\langle u_{s,y} \rangle/u_\eta = -Sv$, with $C$ defined in the text.
	Dashed lines in (b) represent $\langle u_{p,y} \rangle/(\tau_p g) = C{Sv}^{-1}-1$.
	}
	\label{fig:2}
\end{figure}

We then consider the particle vertical mean velocity $\langle u _{p,y}\rangle$.
Since in the considered regimes we have approximately $\langle u_{fp,y} \rangle/u_\eta = C$ and $\langle u_{s,y} \rangle/u_\eta = -Sv$, we expect $\langle u_{p,y} \rangle/u_\eta = C-Sv$.
Figure~\ref{fig:2}a supports this scaling for the ensemble of all particles.
Similarly, normalizing by the still-air terminal velocity leads to $\langle u_{p,y} \rangle/(\tau_p g) = C/{Sv}-1$.
This is confirmed in figure~\ref{fig:2}b, where the scaling is shown to hold also for upward and downward moving particles (with the respective values of the constant).
We again stress that this cannot be extrapolated ad libitum: in the limit of both vanishing and infinite particle inertia, we expect $\langle u_{p,y} \rangle/(\tau_p g) = -1$.

\begin{figure}
	\centering
	\includegraphics{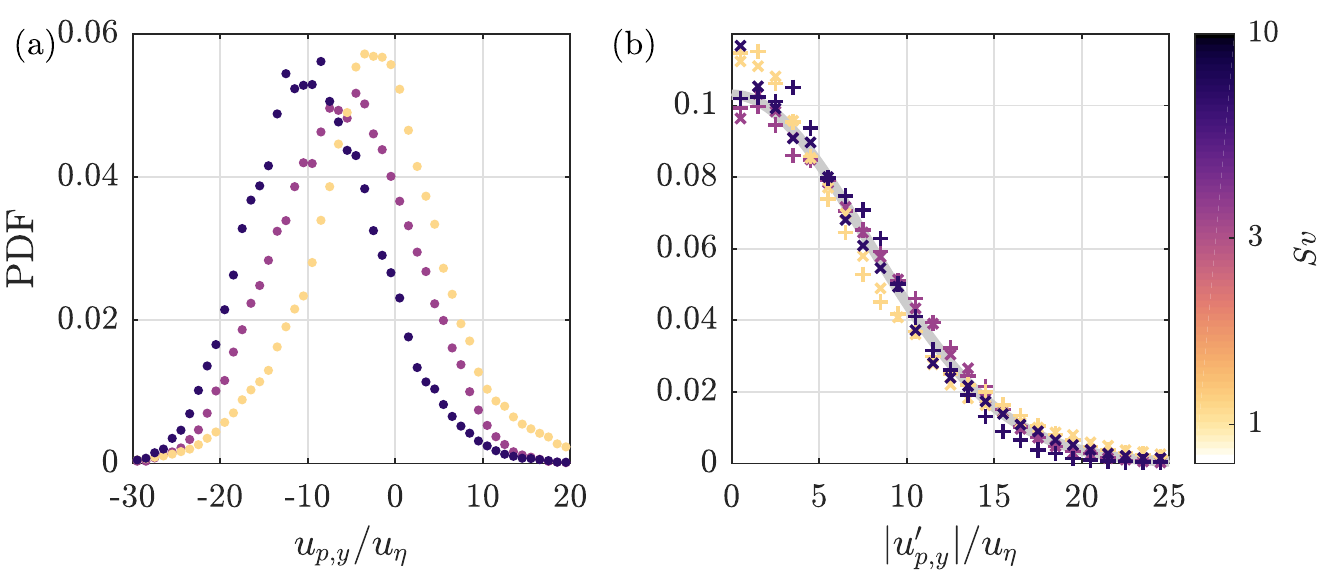}
	\caption{PDFs of the vertical particle velocity for select cases (a).
		PDFs of the absolute value of the fluctuating part of the vertical particle velocity (b). `x' denotes upward moving particles while `+' denotes downward moving particles.
		The  solid grey line indicates a Gaussian distribution.
	}
	\label{fig:3}
\end{figure}

The probability density function (PDF) of the normalized vertical particle velocity, $u_{p,y}/u_\eta$, is shown in figure~\ref{fig:3}a for three selected cases, the other cases sharing the same trends.
For increasing $Sv$, the distributions shifts to more negative values as expected, and the distributions become positively skewed.
This is in contrast with the findings of \citet{Baker2017}, who used point-particle simulations.
The disagreement for the heavier particles is not surprising, as the assumptions behind such simulations become questionable with increasing $Re_p$.
Figure~\ref{fig:3}b reports the PDF of the absolute value of the particle velocity fluctuations, $|u_{p,y}'|/u_\eta$, distinguishing between upward- and downward-moving particles.
We observe no difference between both subsets, and therefore in the following we do not consider them separately.
All $Sv$ cases collapse well on a Gaussian distribution when normalized by the Kolmogorov velocity.
This confirms the dominant role of the fluid fluctuations in determining the particle velocity fluctuations, for a wide range of response times and fall speeds.

\begin{figure}
	\centering
	\includegraphics{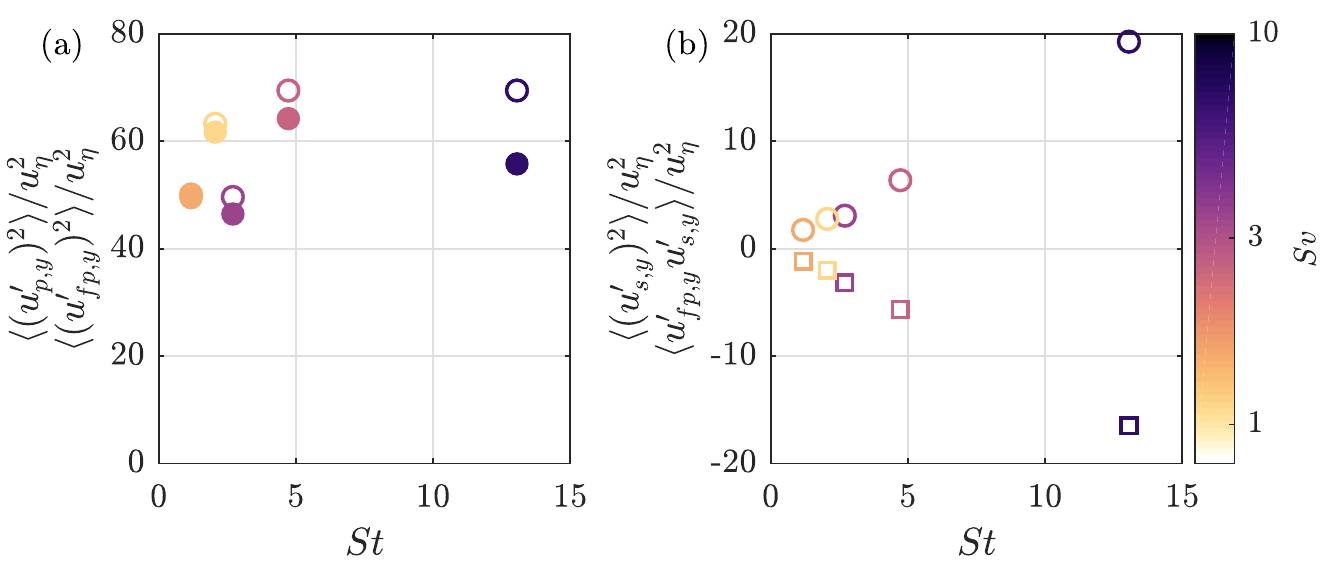}
	\caption{Variance of vertical particle velocity (solid symbols) and sampled-fluid velocity (open symbols) (a).
	Variance of vertical slip velocity (circles) and covariance of the vertical slip velocity and sampled fluid velocity (squares) (b).
	}
	\label{fig:4}
\end{figure}

We then investigate the scaling of the particle velocity fluctuations by considering the variance of $\boldsymbol{u_p}' = \boldsymbol{u_{fp}}' + \boldsymbol{u_{s}}'$:
\begin{equation}
	\langle(\boldsymbol{u_p}')^2\rangle = \langle(\boldsymbol{u_{fp}}')^2\rangle + \langle(\boldsymbol{u_{s}}')^2\rangle + 2 \langle \boldsymbol{u_{fp}}' \boldsymbol{u_{s}}'\rangle
	\label{eqn:up_var}
\end{equation}
Figure~\ref{fig:4} displays the vertical components of the four terms in equation (\ref{eqn:up_var}) for the various cases, normalized by Kolmogorov scaling.
The horizontal components, not shown, behave similarly.
The particle velocity variance $\langle(u_{p,y}')^2\rangle$ is smaller than but comparable to the variance of the sampled-fluid velocity $\langle(u_{fp,y}')^2\rangle$ (figure~\ref{fig:4}a), as also reported by \citet{Ireland2016a} in zero gravity simulations.
This confirms that, in the present range of parameters, the fluctuating energy of the particles is driven by the turbulent kinetic energy.
Figure~\ref{fig:4}b indicates that the normalized slip velocity variance $\langle(u_{s,y}')^2\rangle/u_\eta^2$ varies linearly with $St$.
This is consistent with the scaling $u_s/u_\eta \propto St^{1/2}$ derived by \citet{Balachandar2009} in the present range ($\tau_\eta < \tau_p < T_L$).
The covariance $\langle u_{fp,y}' u_{s,y}'\rangle$ also varies linearly with $St$ and roughly equals $-\langle(u_{s,y}')^2\rangle$.
Therefore, from (\ref{eqn:up_var}) we have $\langle(u_{p,y}')^2\rangle \approx \langle(u_{fp,y}')^2\rangle - \langle(u_{s,y}')^2\rangle$.
As $\langle(u_{s,y}')^2\rangle$ grows with $St$, we retrieve the influence of inertial filtering: heavier particles exhibit weaker velocity fluctuations with respect to the sampled-fluid fluctuations.

\begin{figure}
	\centering
	\includegraphics[]{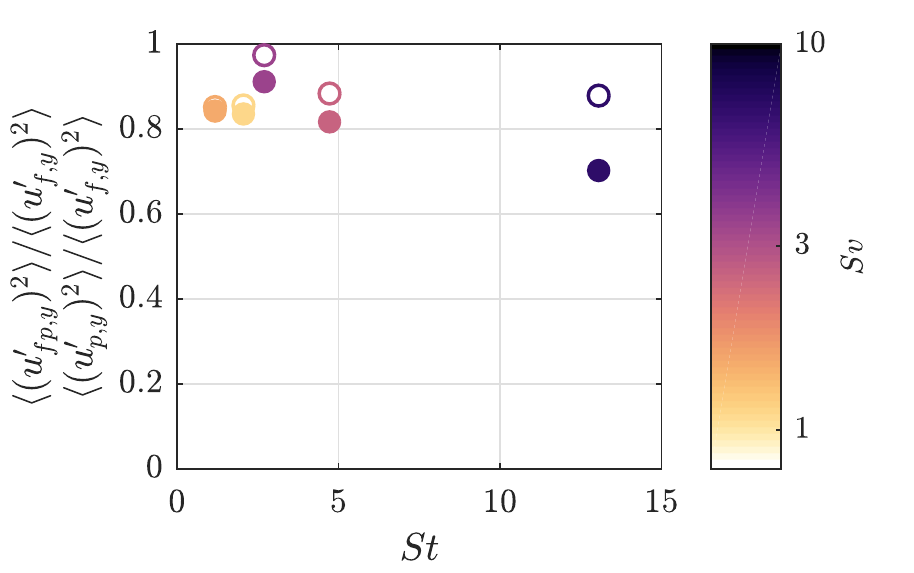}
	\caption{Variances of the vertical sampled-fluid velocity (open symbols) and vertical particle velocity (solid symbols), normalized by the fluid velocity variance.
	}
	\label{fig:5}
\end{figure}

We conclude this section by comparing the sampled-fluid and particle velocity variance against the fluid velocity variance, again focusing on the vertical components (figure~\ref{fig:5}).
This allows us to quantitatively compare the fluctuating energy of the dispersed and carrier phase.
The zero-gravity simulations of \citet{Ireland2016a} indicated that the particle velocity fluctuations can exceed the fluid fluctuations for $St < 1$, due to preferential sampling of energetic flow regions.
In the present case, on the other hand, both inertia and gravity concur to reduce the particle fluctuating energy below the turbulent kinetic energy of the fluid, in agreement with the algebraic model of \citet{Wang1993a} and the results of \citet{Good2014} in similar ranges of $St$ and $Sv$.
In particular, figure~\ref{fig:5} shows that the fluctuating energy of the fluid sampled by the particles is somewhat lower than (although comparable to) the unconditional turbulent kinetic energy.
This implies that, as far as this observable is concerned, the effect of preferential sampling is relatively weak.
The latter is indeed offset by gravitational drift, which reduces the particle ability to follow energetic fluid structures, hence $\langle(u_{fp,y}')^2\rangle < \langle(u_{f,y}')^2\rangle$.
This picture will be confirmed later, when analyzing the fast decorrelation of the sampled-fluid velocity (see \S\ref{ss:timescales}).
Inertial filtering further reduces $\langle(u_{p,y}')^2\rangle$ with respect to $\langle(u_{fp,y}')^2\rangle$, and therefore the former is 10~\% to 30~\% lower than $\langle(u_{f,y}')^2\rangle$.

\subsection{Acceleration}
\begin{figure}
	\centering
	\includegraphics[]{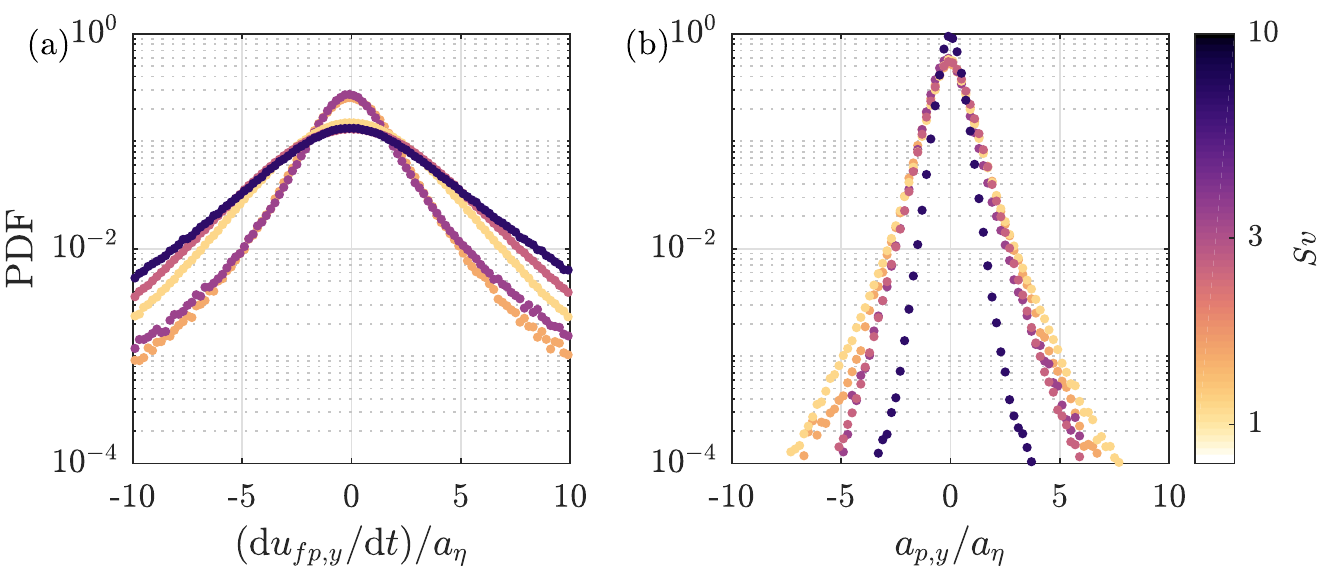}
	\caption{PDFs of the vertical components of the temporal derivative of the sampled-fluid velocity (a) and the particle acceleration (b).
	}
	\label{fig:6}
\end{figure}

We present results for the vertical components of the particle acceleration, $a_{p,y}$, the horizontal components behaving similarly.
In addition, we consider the temporal derivative of the sampled-fluid velocity, $\mathrm{d}u_{fp,y}/\mathrm{d}t$.
The PDFs of the latter are presented in figure~\ref{fig:6}a, showing similar distributions for the cases with the same $Re_\lambda$.
Indeed, for vanishing inertia $\mathrm{d}u_{fp,y}/\mathrm{d}t$ equals the Eulerian acceleration which, in homogeneous turbulence, is a function of $Re_\lambda$ only \citep{Hill2002,Sawford2003}.
For a given $Re_\lambda$, the distributions of $\mathrm{d}u_{fp,y}/\mathrm{d}t$ become wider with $Sv$, in agreement with the simulations of \citet{Ireland2016b}.
This is a manifestation of the crossing-trajectories effect: for higher settling rate, the particles experience rapid changes of the sampled-fluid environment.
The distributions of $a_{p,y}$ (figure~\ref{fig:6}b) are much narrower than the corresponding distributions of $\mathrm{d}u_{fp,y}/\mathrm{d}t$, due to inertial filtering.
This effect becomes stronger with larger particle inertia, which in the present case implies an increase of both $St$ and $Sv$.

\begin{figure}
	\centering
	\includegraphics[]{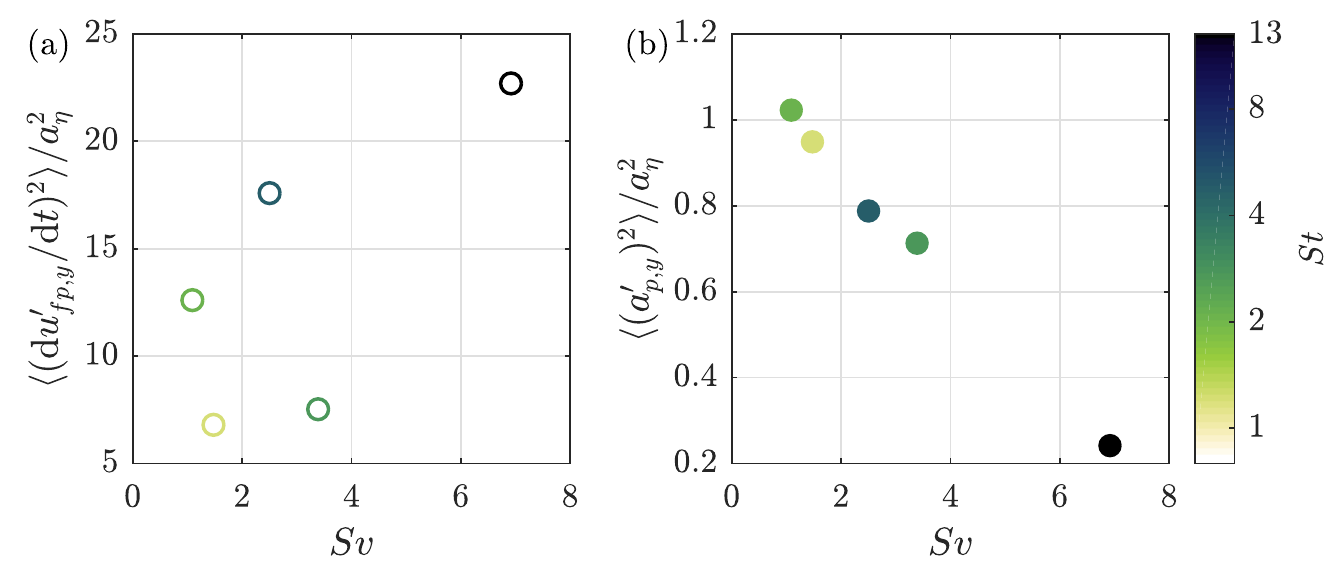}
	\caption{Variances of the vertical components of the temporal derivative of the sampled-fluid velocity (a) and the particle acceleration (b).
	}
	\label{fig:7}
\end{figure}

The variances of $\mathrm{d}u_{fp,y}/\mathrm{d}t$ and $a_{p,y}$ are quantified in figure~\ref{fig:7}.
As observed above, the variance of $\mathrm{d}u_{fp,y}/\mathrm{d}t$ is dominated by $Re_\lambda$ and increases with $Sv$.
For tracers in homogeneous isotropic turbulence, the normalized acceleration variance of the fluid can be approximated as $a_0 \equiv \langle a_f^2 \rangle /a_\eta^2 = 5/(1+100 Re_\lambda^{-1})$ \citep{Sawford2003}, hence for $Re_\lambda = 289$ and 462 we expect $a_0 = 3.6$ and 4.0, respectively.
\citet{Ireland2016b} showed a monotonic increase of $\langle (\mathrm{d}u_{fp,y}'/\mathrm{d}t)^2\rangle/u_\eta^2$ roughly with $Sv^{3/2}$, independent of $Re_\lambda$.
In the present study, the variance of $\mathrm{d}u_{fp,y}/\mathrm{d}t$ is larger than for tracers, but not as large as in \citet{Ireland2016b} and still dependent on $Re_\lambda$.
The difference with \citet{Ireland2016b} is reflected in the Froude number: in their study $Fr=0.052$, while here $Fr=0.8$ and 1.9, for $Re_\lambda = 289$ and 462 respectively.
For $Fr=\textit{O}(1)$, gravitational and Kolmogorov acceleration are of the same order of magnitude, i.e. fluid turbulence and gravity are expected to have comparable influences.
The variance of $a_{p,y}$ (figure~\ref{fig:7}b) decreases approximately linearly with $Sv$ in the present range.
It is significantly smaller than the variance of $\mathrm{d}u_{fp,y}/\mathrm{d}t$ and also smaller than $a_0$, due to inertial filtering. 

\subsection{Preferential sampling}
\begin{figure}
	\centering
	\includegraphics[]{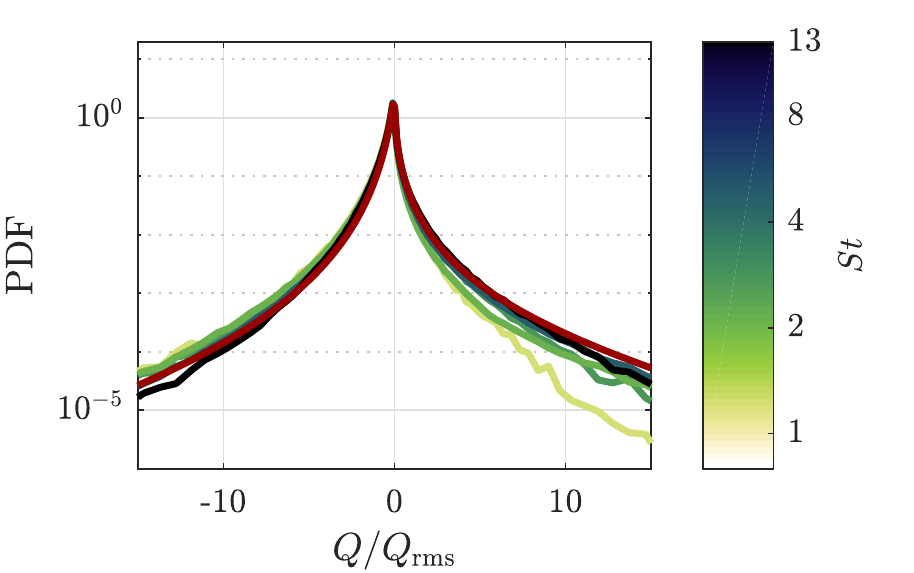}
	\caption{PDFs of the particle-sampled $Q$, defined in (\ref{eqn:Q}). The red line indicates ensemble averaged (i.e. unconditional) $Q$ (similar for both Reynolds numbers).
	}
	\label{fig:8}
\end{figure}

To quantify the extent and influence of preferential sampling, we discriminate between rotation-dominated and strain-dominated fluid regions using the second invariant of the velocity gradient tensor \citep{Hunt1988}:
\begin{equation}
	Q = \omega^2/4 - s^2/2,
	\label{eqn:Q}
\end{equation}
where $\omega^2 = 2 \mathrm{tr}(\boldsymbol{\Omega}^2)$ is the enstrophy, $s^2 = \mathrm{tr}(\bf S^2)$ is the squared strain rate, and $\bf S$ and $\boldsymbol{\Omega}$ are the symmetric and anti-symmetric part of the velocity gradient tensor, respectively.
Due to the planar nature of the measurements, we can only consider the in-plane components of the velocity gradient tensor, which are not sufficient to fully describe the flow topology \citep{Perry1994}.
Still, especially in homogeneous turbulence, the in-plane part of the tensor provides important physical insight into the properties of high-enstrophy and high-strain structures \citep{Cardesa2013} and captures the fundamental small-scale features \citep{Fiscaletti2015,Carter2018}.
Figure~\ref{fig:8} shows the PDF of $Q$ evaluated at the particle locations, compared to the unconditioned fluid.
For the smaller $St$ considered, the particle-conditioned distributions display the expected under-sampling of rotation-dominated regions ($Q>0$), which however becomes progressively weaker as $St$ increases above unity. 

\begin{figure}
	\centering
	\includegraphics[]{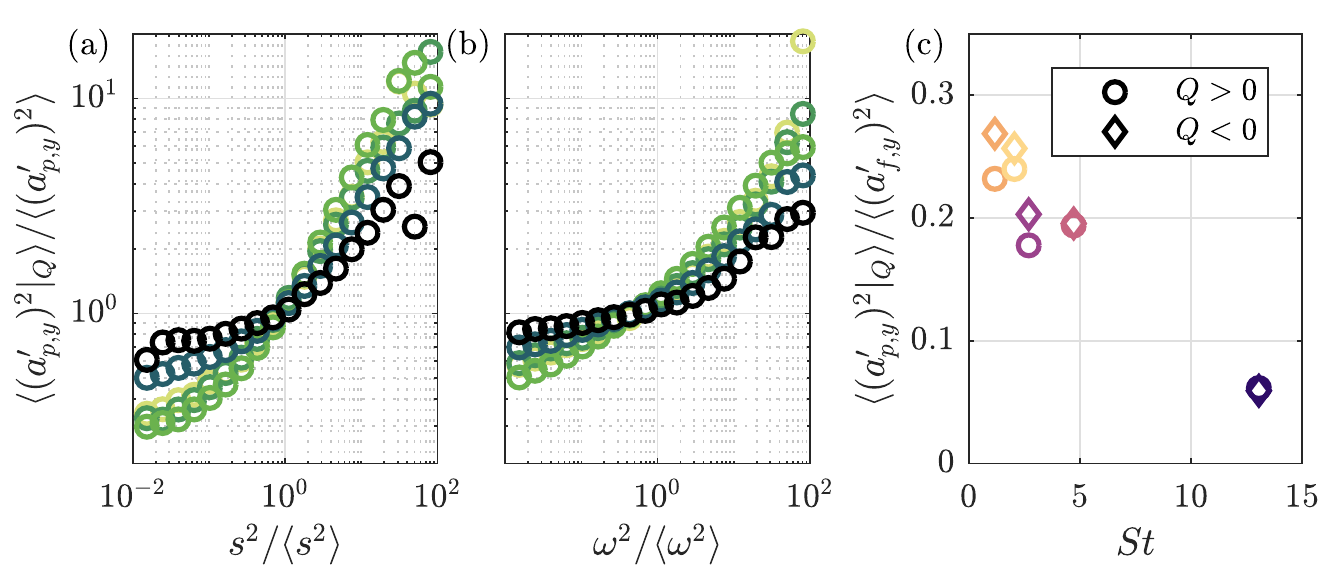}
	\caption{Vertical component of the particle acceleration variance, conditioned to $Q$. Acceleration variance conditioned to high-strain events for $Q<0$ (a) and to high-enstrophy events for $Q>0$ (b), both normalized by the variance of the ensemble.
	Acceleration variance averaged over $Q<0$ (diamonds) and $Q>0$ (circles), normalized by the variance of the fluid acceleration, calculated as $\langle (a_{f,y}')^2\rangle = a_0 a_\eta$.
	}
	\label{fig:9}
\end{figure}

The effect of the small-scale features of the sampled fluid on the particle motion is depicted in figure~\ref{fig:9}, which plots the particle acceleration variance conditioned on strain rate (figure~\ref{fig:9}a) and enstrophy (figure~\ref{fig:9}b).
In both cases, larger levels of small-scale turbulence activity correspond to stronger accelerations.
Although the correlation with $s^2$ is somewhat stronger than with $\omega^2$, the similarity of both plots is consistent with the view that high-strain and high-enstrophy events are often concurrent \citep{Worth2011,Yeung2012,Carter2018}.
Due to inertial filtering, the impact of the sampled-fluid topology decreases steeply with $St$.
This is clearly shown in figure~\ref{fig:9}c, displaying the variance of the particle acceleration conditioned on the sign of $Q$: particles in strain-dominated regions do display larger accelerations \citep{Bec2006}, but the effect becomes unmeasurable for $St = 4.7$ and larger.

\subsection{Structure functions and pair dispersion}
In this section we consider two-particle statistics, starting with the 2nd-order Eulerian velocity structure function $\boldsymbol{S_2}(\boldsymbol{r}) = \langle ( \boldsymbol{u}'(\boldsymbol{x}) - \boldsymbol{u}'(\boldsymbol{x}+\boldsymbol{r}))^2\rangle$, where $\boldsymbol{r}$ is the separation vector.
For fluid tracers in homogeneous isotropic turbulence, this scales as $r^2$ in the dissipative range, plateaus to the fluid velocity variance at large-scale separations, and follows the scaling predicted by \citet{Kolmogorov1941} in the inertial range:
\begin{equation}
	S_{2||}(r) = C_2(\varepsilon r)^{2/3}	\hspace{2em} \mathrm{or} \hspace{2em}  S_{2||}/u_\eta^2 = C_2(r/\eta)^{2/3},
\end{equation}
\begin{equation}
S_{2\perp}(r) = \frac{4}{3} C_2(\varepsilon r)^{2/3} \hspace{2em} \mathrm{or} \hspace{2em}  S_{2\perp}/u_\eta^2 = \frac{4}{3}C_2(r/\eta)^{2/3},
\end{equation}
where the subscripts $||$ and $\perp$ denote velocity components longitudinal and transverse to the separation vector, respectively, and $C_2\approx 2$ \citep{Saddoughi1994}.
For non-tracer particles, inertia and gravity modify these trends.
Simulations and experiments indicate that, when gravitational effects are negligible or absent, inertia leads to greater relative velocities between nearby particles, such that the structure function increasingly deviates from the $r^2$ scaling at small separations \citep{Bec2010,Ireland2016a,Dou2018}.
This is attributed to the path-history effect, i.e. the particles retaining memory of their past interactions with the flow and thus approaching each other with a significant uncorrelated velocity component (see, among many others, \citealt{Wilkinson2005,Fevrier2005,Bragg2014b,Fong2019}).
With the addition of gravity, the simulations of \citet{Ireland2016b} indicated a strong reduction of relative particle velocity at all separations.
They attributed this to the decorrelation of the sampled-fluid velocity along the particle trajectories (which we shall confirm later), hindering the path-history effect and in turn decreasing the relative velocities.
To our best knowledge, no previous experimental observation could verify this latter point.

\begin{figure}
	\centering
	\includegraphics[]{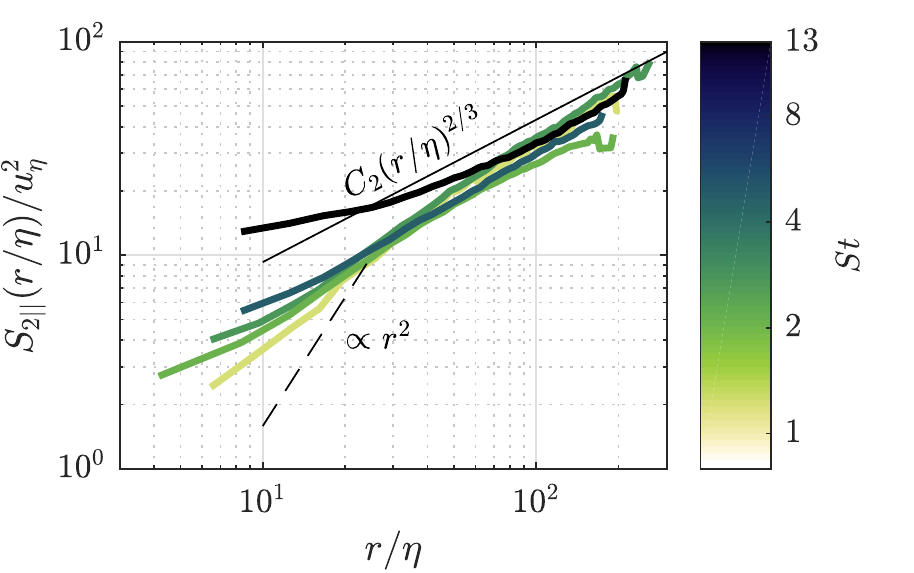}
	\caption{Longitudinal structure functions using the particle velocity. The dashed line indicates $r^2$ scaling. The solid line indicates the prediction by \citet{Kolmogorov1941} for the inertial range.
	}
	\label{fig:10}
\end{figure}

The longitudinal structure functions are presented in figure~\ref{fig:10}, the transverse components (not shown) showing analogous trends.
Due to the finite laser sheet thickness ($\approx 6 \eta$) we expect an overestimation of the relative velocity over the dissipative range \citep{Dou2018}, which however may not overwhelm the trend.
At small separations ($r\lesssim20\eta$), the structure functions deviate from the $r^2$ scaling with increasing $St$, confirming previous findings.
In the inertial range, $S_{2||}(r)$ roughly follows the $r^{2/3}$ scaling, but the values are significantly lower than the \citet{Kolmogorov1941} expectation for tracers (which the turbulence in our chamber closely approximates, \citet{Carter2017}).
The gap persists at large scales, consistent with the fact that the inertial particle fluctuating energy (to which the structure function asymptotes for large $r$) is lower compared to the fluid, see figure~\ref{fig:5}.
Although the competing effects of inertia and gravity cannot be separated here, these results appear to confirm the observation of \citet{Ireland2016b} that gravity reduces the relative particle velocities at all scales.

We then turn to particle pair separations as a function of time, $r(t)$.
In homogeneous isotropic turbulence, for tracer pairs with an initial separation $r_0$ in the inertial range, we expect the mean square separation to follow the ballistic scaling proposed by \citet{Batchelor1950}:
\begin{equation}
	\langle (r(t) - r_0)^2 \rangle = \frac{11}{3} C_2 (\varepsilon r_0)^{2/3} t^2 \hspace{3em} \mathrm{for} \hspace{1em} \tau_\eta \ll t \ll t_B
	\label{eqn:pairdisp}
\end{equation}
where $t_B = (r_0^2/\varepsilon)^{1/3}$ is the characteristic time scale of an eddy of size $r_0$.
For $t_B \ll t \ll T_L$, the dispersion does not depend on $r_0$ and is expected to follow the Richardson-Obukhov scaling $\langle r(t)^2\rangle = g \varepsilon t^3$, where $g \approx 0.5$ \citep{Salazar2012}.
For $t$ in the dissipative range, particle inertia enhances pair dispersion at small times, due to the large relative velocities at small separation \citep{Bec2010,Gibert2010}; while for larger $t$, the inertial filtering and path-history effects reduce pair dispersion compared to tracers \citep{Bragg2016}.
To our best knowledge, the only previous investigations on the effect of gravity on pair dispersion are the numerical studies by \citet{Chang2015} and \citet{Dhariwal2019}, who mostly focused on bi-dispersed particles sets. 

\begin{figure}
	\centering
	\includegraphics[scale=0.82]{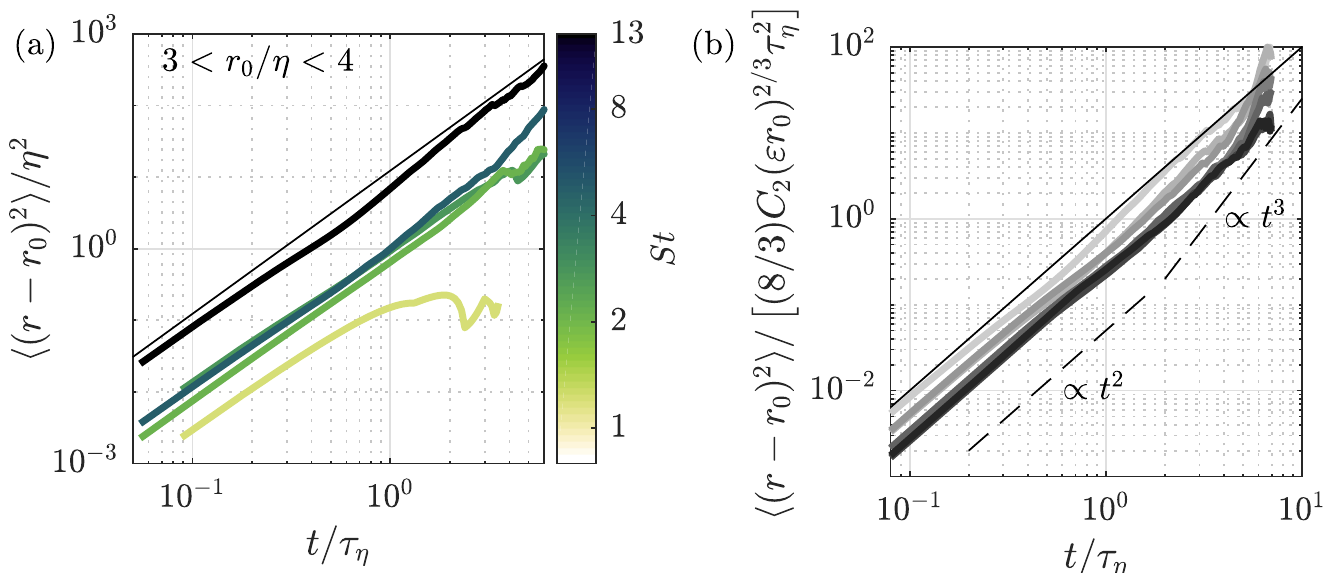}
	\caption{Mean square separation of particle pairs with initial separation $r_0$. Mean square separation of pairs with $3<r_0/\eta<4$ for all cases (a). Mean square separation of pairs with $3<r_0/\eta<4$ to $9<r_0/\eta<10$ (increasing with darker shades) for the $St=13$ case. The solid thin lines indicates the scaling in (\ref{eqn:pairdisp}).
	}
	\label{fig:11}
\end{figure}

The mean square separation for our inertial particles is presented in figure~\ref{fig:11}.
Due to the nature of the measurements, the results are biased by the constraint that the trajectories cannot separate more than the laser sheet thickness in $z$.
One way to account for this is to consider that the right-hand-side in (\ref{eqn:pairdisp}) is the geometric average of the structure function components at $r_0$, i.e. $\langle (r(t) - r_0)^2 \rangle = \left(S_{2||}(r_0) + 2 S_{2\perp}(r_0)\right)t^2$.
Setting to zero the out-of-plane velocity, we can write the mean square separation for tracers as $\frac{8}{3} C_2 (\varepsilon r_0)^{2/3} t^2$, which is plotted as a black solid line for reference.
The effect of $St$ (and indirectly of $Sv$) is represented in figure~\ref{fig:11}a, for trajectories with initial separation $3<r_0/\eta<4$.
With increasing particle inertia, the mean square separation grows.
Still, all curves are generally below the expectation for tracers, illustrating the competing effect of inertia and gravity.
The normalized form plotted in figure~\ref{fig:11}b illustrates the effect of the initial separation for the case $St=13$: with increasing initial separations, the impact of the large relative velocities is weakened, and the curves tend to collapse on each other (which is also the case for smaller $St$, not shown).
This confirms the strong influence of the uncorrelated motion of near-by particles, already highlighted by the structure functions.
The plot also emphasizes how the particles follow the Batchelor regime up to $t=\textit{O}(\tau_\eta)$, after which the slope increases.
However, given the bias due to the shape of the illuminated volume, caution should be exerted when interpreting the behavior for relatively long times.
Here we just note that such an early transition out of the ballistic regime (which for tracers is expected to ensue at later times, \citealt{Bitane2012}) could be associated to the time scale of eddy-crossing by the falling particles, similarly to recent findings for rising bubbles \citep{Mathai2018}.
Further research is warranted on this point. 

\section{Modeling the velocity and acceleration variances} \label{s:model}
Here we leverage and expand on the framework of \citet{Csanady1963} to obtain expressions for the particle velocity and acceleration variances.
To summarize the detailed explanation that follows, we integrate the Lagrangian particle velocity spectrum, which is obtained from the sampled-fluid velocity spectrum modulated by a response function.
To extend the frequency range of the spectrum, we Fourier-transform the \citet{Sawford1991} expression of the velocity autocorrelation.
As this was originally derived for the unconditional fluid, we substitute in it the Lagrangian time scales associated to the sampled fluid.

\subsection{Lagrangian spectrum of particle velocity}
For frequencies in the inertial range $\upi/T_L \ll \omega \ll \upi/\tau_\eta$, the Lagrangian spectrum for the fluid flow is expected to scale as $E\propto \varepsilon \omega^{-2}$ \citep{Tennekes1972,Yeung2001}.
The spectrum can also be derived from the velocity autocorrelation as the two form a Fourier transform pair.
For the canonical form of the autocorrelation, $R(t) = \langle (u')^2 \rangle \mathrm{exp}(-t/T_L)$, one has $E = \langle (u')^2 \rangle T_L / (1+(\omega T_L)^2)$ \citep{Hinze1975,Mordant2001,Zhang2019}, valid for $\omega \ll \upi/\tau_\eta$ (or $\tau_\eta \ll t$).
For small time separations, the autocorrelation deviates from the exponential, tending to a horizontal asymptote at $t=0$ with a curvature proportional to the acceleration variance \citep{Mordant2004}.
\citet{Sawford1991} proposed an alternative formulation valid for all $t$, using a second timescale $T_2$ related to the finite acceleration variance:
\begin{equation}
	R(t) = \langle (u')^2 \rangle \frac{T_L \mathrm{exp}(-t/T_L) - T_2 \mathrm{exp}(-t/T_2)}{T_L - T_2},
	\label{eqn:autocorr}
\end{equation}
Fourier transformation yields:
\begin{equation}
	E(\omega) = \frac{\langle (u')^2 \rangle}{T_L - T_2} \left[ \frac{T_L^2}{1+(\omega T_L)^2} - \frac{T_2^2}{1+(\omega T_2)^2}\right]
\end{equation}
We use this expression to model the Lagrangian spectrum of the sampled-fluid velocity, by substituting the respective time scales $T_{L,fp}$ and $T_{2,fp}$.
Using the response function (\ref{eqn:response}) in combination with (\ref{eqn:uspectra}) and (\ref{eqn:aspectra}), we have:
\begin{equation}
	\langle(\boldsymbol{u_p}')^2\rangle = \langle(\boldsymbol{u_{fp}}')^2\rangle \left[1- \frac{St^2}{(T_{L,fp}/\tau_\eta+St)(T_{2,fp}/\tau_\eta+St)}\right],
	\label{eqn:umodel}
\end{equation}
\begin{equation}
\langle(\boldsymbol{a_p}')^2\rangle = \tau_\eta^{-2} \langle(\boldsymbol{u_{fp}}')^2\rangle \frac{1}{(T_{L,fp}/\tau_\eta+St)(T_{2,fp}/\tau_\eta+St)}.
\label{eqn:amodel}
\end{equation}

\subsection{Time scales of the sampled-fluid velocity}\label{ss:timescales}
Because evaluating equations (\ref{eqn:umodel}) and (\ref{eqn:amodel}) requires estimates for $T_{L,fp}$ and $T_{2,fp}$, we consider the issue of how those compare to $T_L$ and $T_2$.
For tracers, the Lagrangian integral timescale is related to the Eulerian integral timescale ($T_E$) as \citep{Yeung2001}:
\begin{equation}
	T_L/T_E \approx 5/C_0
\end{equation}
where $C_0$ is the pre-factor in the expression of the Lagrangian velocity structure function.
This depends on the turbulence Reynolds number, and based on a review of the literature \citet{Lien2002} suggested $C_0 = C_0^\infty (1-(0.1 Re_\lambda)^{1/2})$ with $C_0^\infty = 6 \: \pm \: 0.5$ \citep{Ouellette2006}.
Considering inertial particles, the Lagrangian integral time scale of the sampled fluid $T_{L,fp}$ is influenced by both inertia and gravity.
For the range of $St$ in the present study, however, the effect of $St$ as derived empirically by \citet{Wang1993a} is negligible.
(Alternatively, \citet{Jung2008} proposed an empirical expression of the $St$ effect derived in zero-gravity.)
Gravity decreases $T_{L,fp}$ due to the crossing-trajectories effect, for which we adopt the expression proposed by \citet{Csanady1963} (see also \citealt{Pozorski1998}):
\begin{equation}
	T_{L,fp} = T_L \frac{1}{\left[1 + \alpha (T_L/T_E)^2 Sv^2 (u_\eta/u_\mathrm{rms})^2\right]^{1/2}}
	\label{eqn:TLfp}
\end{equation}
where $\alpha = 1$ and 4 for the time scales associated to the vertical and horizontal components, respectively. Alternatively, using $\langle (u_f')^2 \rangle/u_\eta^2 = Re_\lambda/\sqrt{15}$ \citep{Hinze1975} and $T_L/\tau_\eta = 2(Re_\lambda+32)/(\sqrt{15}C_0)$ \citep{Zaichik2003}, this can be expressed as
\begin{equation}
	\frac{T_{L,fp}}{\tau_\eta} = \frac{2 (Re_\lambda+32)}{\sqrt{15} C_0} \frac{1}{\left[1 + \alpha (5/C_0)^2 Sv^2 (\sqrt{15}/Re_\lambda)^2\right]^{1/2}},
	\label{eqn:TLfp2}
\end{equation}
which is a function of $Re_\lambda$ and $Sv$ only.

\begin{figure}
	\centering
	\includegraphics[]{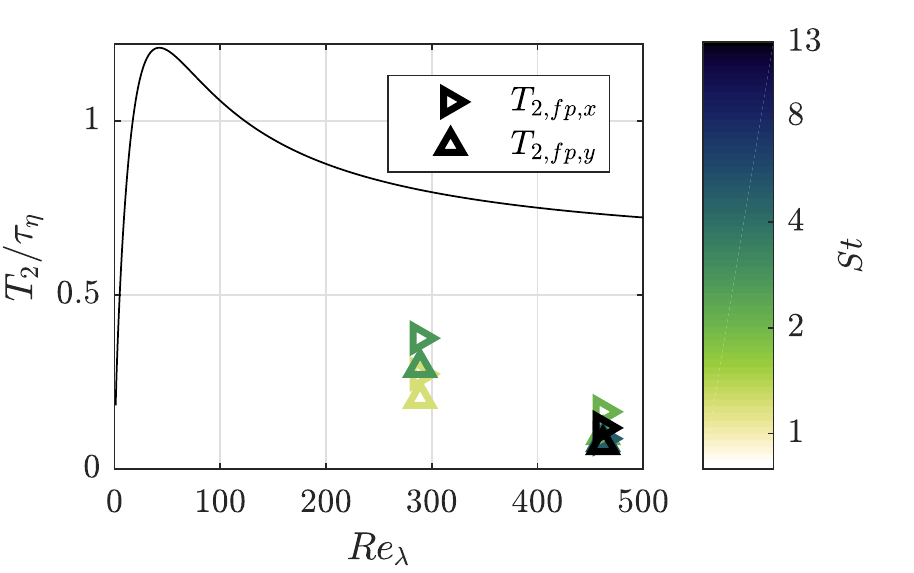}
	\caption{Empirical values of $T_{2,fp}$, determined using (\ref{eqn:T2emp}). The black line indicates the prediction for tracers in (\ref{eqn:T2}).
	}
	\label{fig:12}
\end{figure}

For tracers, the time scale $T_2$ can be written as \citep{Sawford1991}:
\begin{equation}
	T_2/\tau_\eta = C_0/(2a_0).
	\label{eqn:T2}
\end{equation}
A relation for $T_{2,fp}$ is obtained from the variance of $\mathrm{d}u_{fp,y}/\mathrm{d}t$:
\begin{equation}
	\langle \left(\frac{\mathrm{d}\boldsymbol{u_{fp,y}}'}{\mathrm{d}t}\right)^2 \rangle = \frac{2}{\upi} \int_0^\infty \omega^2 \boldsymbol{E_{fp}}(\omega) \mathrm{d}\omega = \frac{\langle (\boldsymbol{u_{fp}}')^2\rangle}{T_{L,fp} T_{2,fp}}
	\label{eqn:T2emp}
\end{equation}
Thus, we can empirically evaluate $T_{2,fp}$ using the measured values for $\langle \left(\mathrm{d}\boldsymbol{u_{fp}'}/\mathrm{d}t\right)^2 \rangle$ and $\langle (\boldsymbol{u_{fp}}')^2\rangle$.
The values are presented in figure~\ref{fig:12}, along with the prediction for tracers in (\ref{eqn:T2}).
$T_{2,fp}$ is measurably smaller than $T_2$, though no clear trend with $St$ is discerned.

\begin{figure}
	\centering
	\includegraphics[]{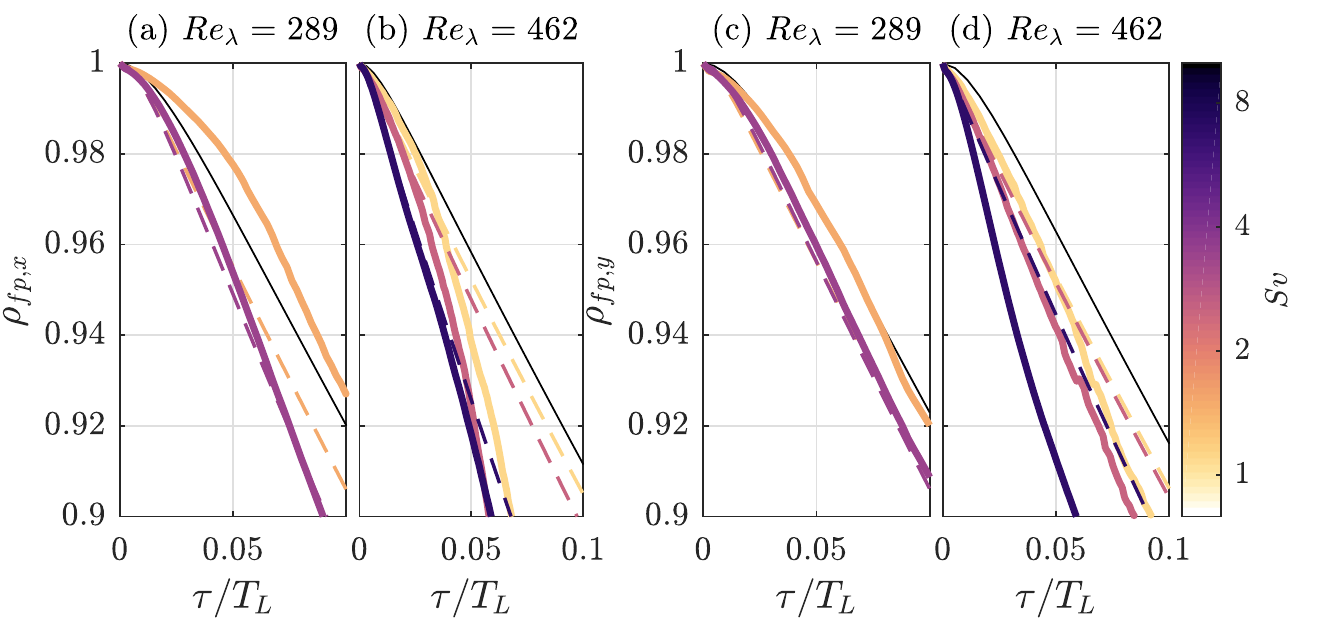}
	\caption{Autocorrelation of the sampled-fluid velocity. Horizontal components for the $Re_\lambda=289$ (a) and 462 (b) cases; vertical components for the $Re_\lambda=289$ (c) and 462 (d) cases.
	Thick solid lines indicate the measured autocorrelation. Thin black lines indicate the prediction for tracers in (\ref{eqn:autocorr}). Dashed lines indicate the prediction from (\ref{eqn:autocorr}) with corrected timescales for inertial particles using (\ref{eqn:TLfp2}) and (\ref{eqn:T2emp}).
	}
	\label{fig:13}
\end{figure}

\begin{figure}
	\centering
	\includegraphics[]{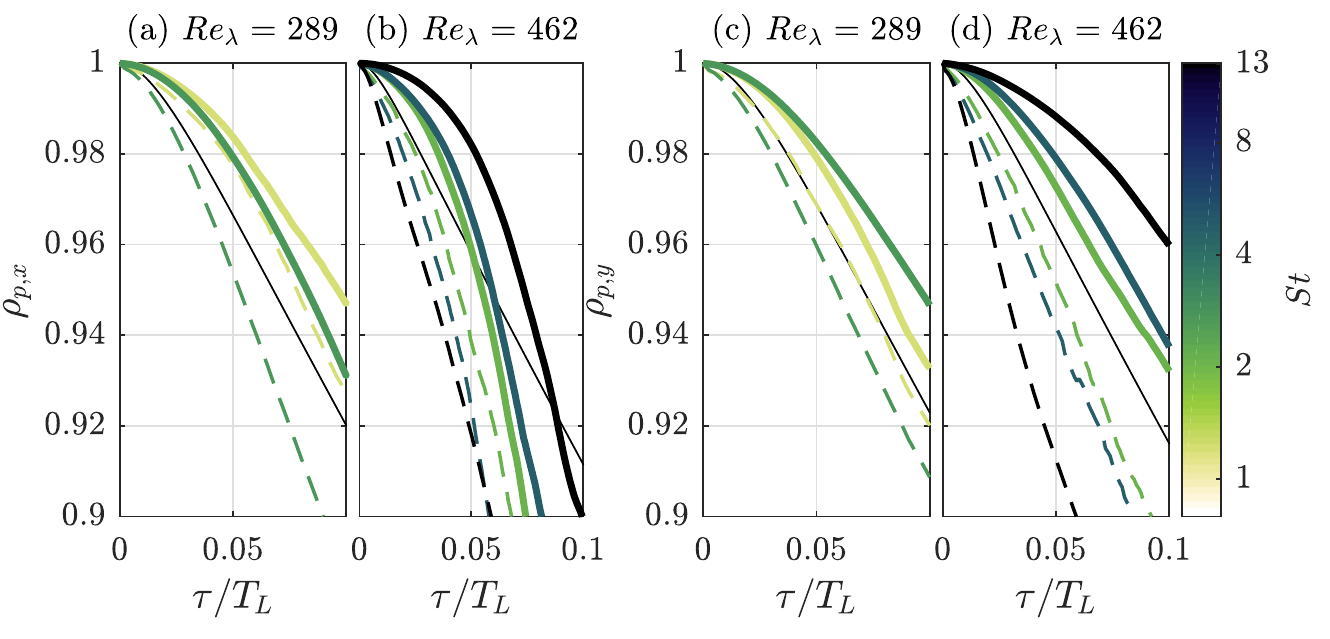}
	\caption{Autocorrelation of the particle velocity. Horizontal components for the $Re_\lambda=289$ (a) and 462 (b) cases; vertical components for the $Re_\lambda=289$ (c) and 462 (d) cases.
	Thick solid lines indicate the autocorrelation. Thin black lines indicate the prediction for tracers in (\ref{eqn:autocorr}). Dashed lines indicate the autocorrelation of the particle-sampled fluid velocity as presented in figure~\ref{fig:13}.
	}
	\label{fig:14}
\end{figure}

With the above estimates of $T_{L,fp}$ and $T_{2,fp}$, we use (\ref{eqn:autocorr}) to model the autocorrelations of the sampled-fluid velocity, $R_{fp}$.
We consider both horizontal and vertical components, and present the results as correlation coefficients, $\rho_{fp}$, normalizing by the variance of the sampled-fluid velocity for the respective components.
The modeled curves are plotted in figure~\ref{fig:13}, along with the corresponding measurements and the fluid velocity autocorrelation (which is also modeled via (\ref{eqn:autocorr}), using $T_L$ and $T_2$).
For clarity we separate both $Re_\lambda$ cases, as these have different fluid time scales.
The plots only extend to values of $t$ for which the measured autocorrelations are based on at least 100 trajectories.
This allows for limited time lags, still sufficient to highlight the trends.
The sampled-fluid velocity decorrelates faster with $Sv$ due to the crossing-trajectory effect.
Despite quantitative differences with the measurements, (\ref{eqn:autocorr}) captures well this trend for short times.
In figure\ref{fig:14} we compare the measured $\rho_{fp}$ against the particle velocity autocorrelation coefficients, $\rho_p$; the fluid velocity autocorrelation is also included for reference.
The apparent trend is that $\rho_p$ decays more slowly for heavier and heavier particles, which is a consequence of inertia; while $\rho_{fp}$ decays more rapidly, which is a consequence of gravity.

\subsection{Model predictions}
\begin{figure}
	\centering
	\includegraphics[]{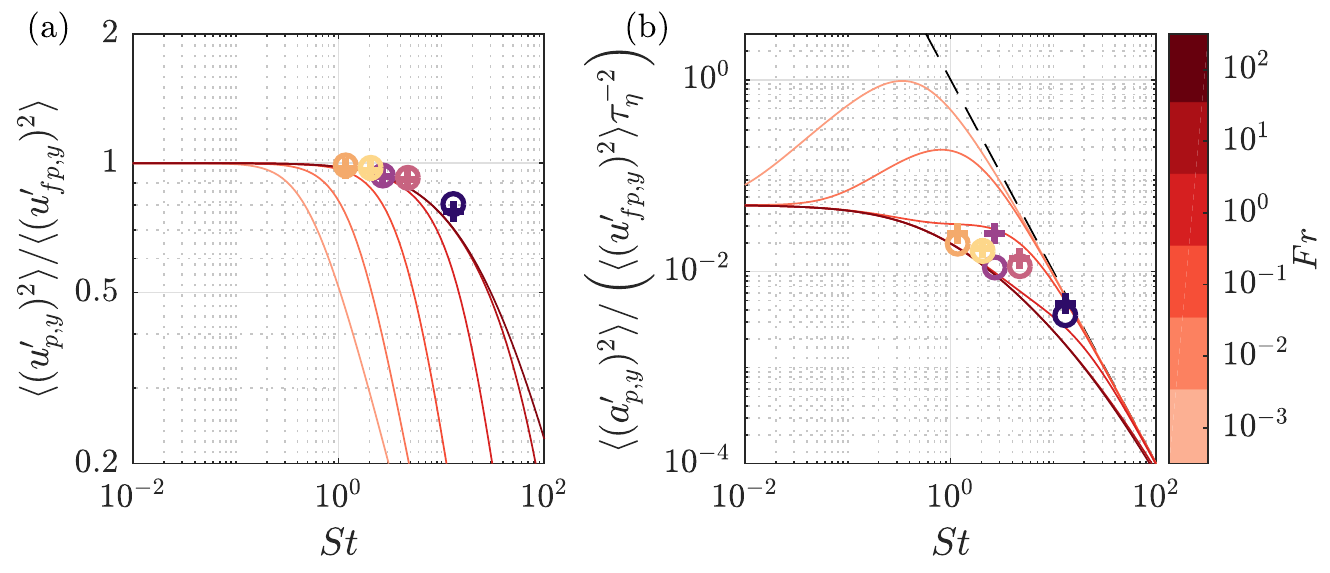}
	\caption{Modeled variances of the vertical component of the particle velocity (a) and acceleration (b).
		Red lines indicate the predictions from (\ref{eqn:umodel2}) and (\ref{eqn:amodel2})), coloured by $Fr$. Circles indicated measured data, coloured by $Sv$ (see, for example figure~\ref{fig:13} for colour coding). `+' signs indicate case-specific predictions as discussed in the text for matching colours.
		The dashed line in (b) indicates the asymptote ${St}^{-2}$.
	}
	\label{fig:15}
\end{figure}

We now compare the measured variances of the particle velocity and particle acceleration against the respective predictions from equations (\ref{eqn:umodel}) and (\ref{eqn:amodel}).
In non-dimensional form:
\begin{equation}
	\frac{\langle(\boldsymbol{u_p}')^2\rangle}{\langle(\boldsymbol{u_{fp}}')^2\rangle} = 1- \frac{St^2}{(T_{L,fp}/\tau_\eta+St)(T_{2,fp}/\tau_\eta+St)},
	\label{eqn:umodel2}
\end{equation}
\begin{equation}
	\frac{\langle(\boldsymbol{a_p}')^2\rangle}{\langle(\boldsymbol{u_{fp}}')^2\rangle \tau_\eta^{-2}} =  \frac{1}{(T_{L,fp}/\tau_\eta+St)(T_{2,fp}/\tau_\eta+St)},
	\label{eqn:amodel2}
\end{equation}
These are shown in figure~\ref{fig:15}a and~b, respectively, as a function of $St$.
The expressions are essentially model forms of the particle response function, as they describe the variance of the particle velocity and acceleration with respect to the sampled-fluid flow.
Here, and in the following, we again consider the vertical components only, the horizontal components leading to analogous conclusions.
These ``generalized predictions'' are compared with the measurements and with ``case-specific predictions''.
The latter differ from the generalized predictions due to the difference in $T_{2,fp}$: for the generalized model we approximate $T_{2,fp} \approx T_2$ (which is calculated for any $St$ according to (\ref{eqn:T2})), whereas for the case-specific predictions we use the empirically determined values for $T_{2,fp}$ (figure~\ref{fig:12}).
As can be seen, this simplification has a moderate impact and does not alter the trends.
In the small $St$ limit, the modeled particle velocity variance equals the variance of the sampled-fluid velocity, regardless of $Fr$.
For finite $St$, the particle velocity variance $\langle(u_p')^2\rangle/\langle(u_{fp}')^2\rangle$ decreases with both inertia and gravity (figure~\ref{fig:15}a).
The dependence with $St$ reflects inertial filtering, while the effect of $Fr$ reflects the decrease of $T_{L,fp}$ with gravitational drift, which in turn damps $\langle(u_p')^2\rangle$ (\ref{eqn:umodel2}).
The effect of gravity becomes negligible for $Fr \gtrsim 10$.
In the limits of either strong gravity ($Fr \ll 1$) or large inertia ($St\gg 1$), $\langle(u_p')^2\rangle$ vanishes compared to $\langle(u_{fp}')^2\rangle$.
Overall, the generalized model represents well the present data. 

The normalized particle acceleration variance $\langle(a_p')^2\rangle/(\langle(u_{fp}')^2\rangle \tau_\eta^{-2})$ (figure~\ref{fig:15}b) tends to $a_0/u_0$ for small $St$, where $u_0 \equiv \langle a_f^2\rangle/u_\eta^2$ is a function of $Re_\lambda$.
For large $St$, it asymptotes to $St^{-2}$, or $\langle(a_p')^2\rangle = \langle(u_{fp}')^2\rangle \tau_p^{-2}$.
For $Fr \ll 1$ the particle acceleration variance displays a non-monotonic behavior with $St$, due to the competing effects of inertia and gravity: an increase of inertia for a fixed $Fr$ implies an increase of $Sv$ and therefore of crossing-trajectory drift, augmenting $\mathrm{d}u_{fp,y}/\mathrm{d}t$ and in turn also the particle acceleration.
As $St$ increases further, inertial filtering eventually dominates and the particle acceleration is dampened.

\begin{figure}
\centering
\includegraphics[]{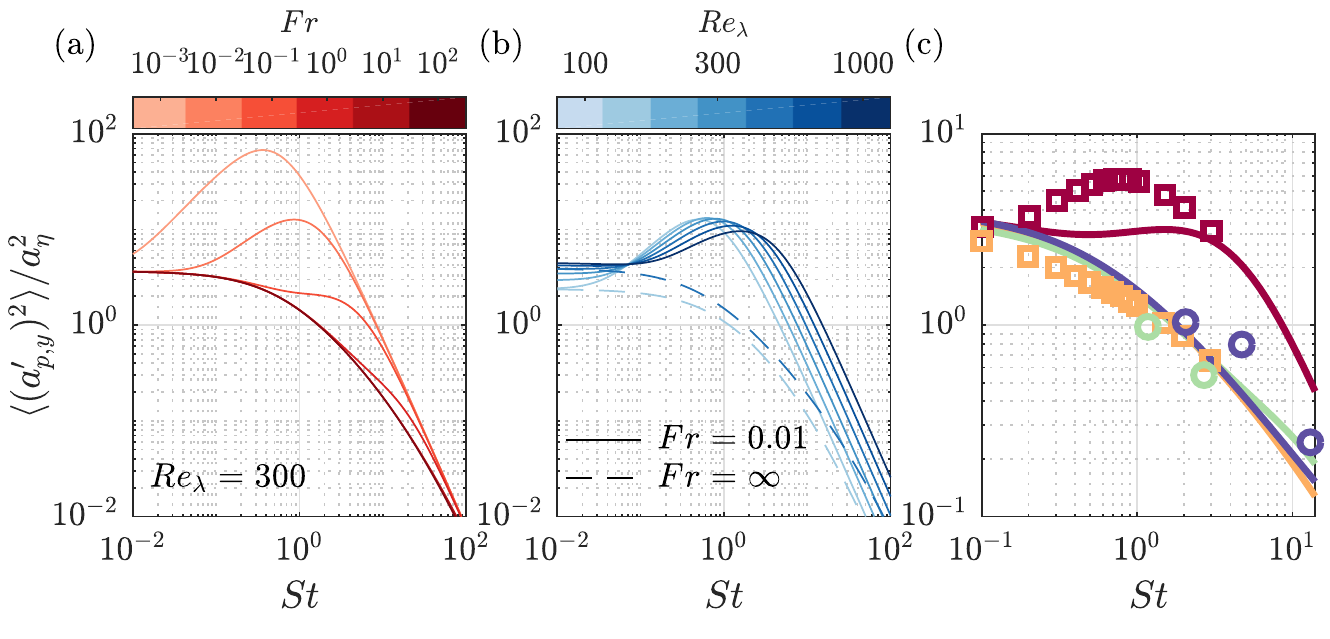}
\caption{Vertical component of the particle acceleration variance, modeled using (\ref{eqn:amodel3}). Variation in $Fr$ for $Re_\lambda = 300$ (a). Variation in $Re_\lambda$ for $Fr = 0.01$ (solid lines) and $Fr=\infty$ (dashed lines) (b). Comparison of cases from \citet{Ireland2016b} (squares) and the present study (circles) with the prediction in (\ref{eqn:amodel3}) (lines) (c). Matching colours have matched $Fr$ and $Re_\lambda$.
}
\label{fig:16}
\end{figure}

As the sampled-fluid properties are not known a priori, (\ref{eqn:umodel2}) and (\ref{eqn:amodel2}) have limited predictive power.
However, as shown in figure~\ref{fig:5}, the sampled-fluid velocity variance is marginally smaller than the fluid velocity variance, and one can approximate $\langle (u_f')^2 \rangle \approx \langle (u_{fp}')^2 \rangle$.
Taking $\langle (u_f')^2 \rangle/u_\eta^2 = Re_\lambda/\sqrt{15}$ \citep{Hinze1975} and substituting in (\ref{eqn:amodel2}) leads to:
\begin{equation}
	\frac{\langle(\boldsymbol{a_p}')^2\rangle}{a_\eta^2} = \frac{Re_\lambda}{\sqrt{15}} \frac{1}{(T_{L,fp}/\tau_\eta+St)(T_2/\tau_\eta+St)}
	\label{eqn:amodel3}
\end{equation}
where we assumed $T_{2,fp} \approx T_2$ and $T_{L,fp}$ is obtained from (\ref{eqn:TLfp}).
Because of the assumptions made, (\ref{eqn:amodel3}) is expected to underpredict somewhat the observations.
The dependence on $St$, $Fr$ and $Re_\lambda$ is illustrated in figure~\ref{fig:16}.
For large $St$, the curves asymptote to $Re_\lambda St^{-2}/\sqrt{15}$.
At intermediate $St$ and $Fr \ll 1$, we retrieve the non-monotonic behavior discussed above.
This non-monotonicity was also reported by \citet{Ireland2016b} for $Fr = 0.052$.
In general (\ref{eqn:amodel3}) agrees remarkably well with the trends reported by their simulations as well as with our experimental data.
The present model demonstrates how the value of $St$ that maximizes the particle acceleration increases with $Fr$, while the maximum shrinks and eventually disappears for $Fr \gtrsim 0.1$.
This is the reason why the non-monotonic behavior found by \citet{Ireland2016b} at $Fr = 0.052$ is not seen in our study at $Fr = \textit{O}(1)$, nor in previous studies at $Fr\approx 0.3$ \citep{Ayyalasomayajula2006} and $Fr \approx 30$ \citep{Volk2008}.
For increasing $Re_\lambda$, the maximum of $\langle(a_p')^2\rangle/a_\eta^2$ is found at larger $St$.
This is consistent with the view that, with increasing $Re_\lambda$, a wider range of scales may influence the particle motion, and therefore also particles with $St > \textit{O}(1)$ may respond strongly to the turbulence \citep{Yoshimoto2007}.

\begin{figure}
	\centering
	\includegraphics[]{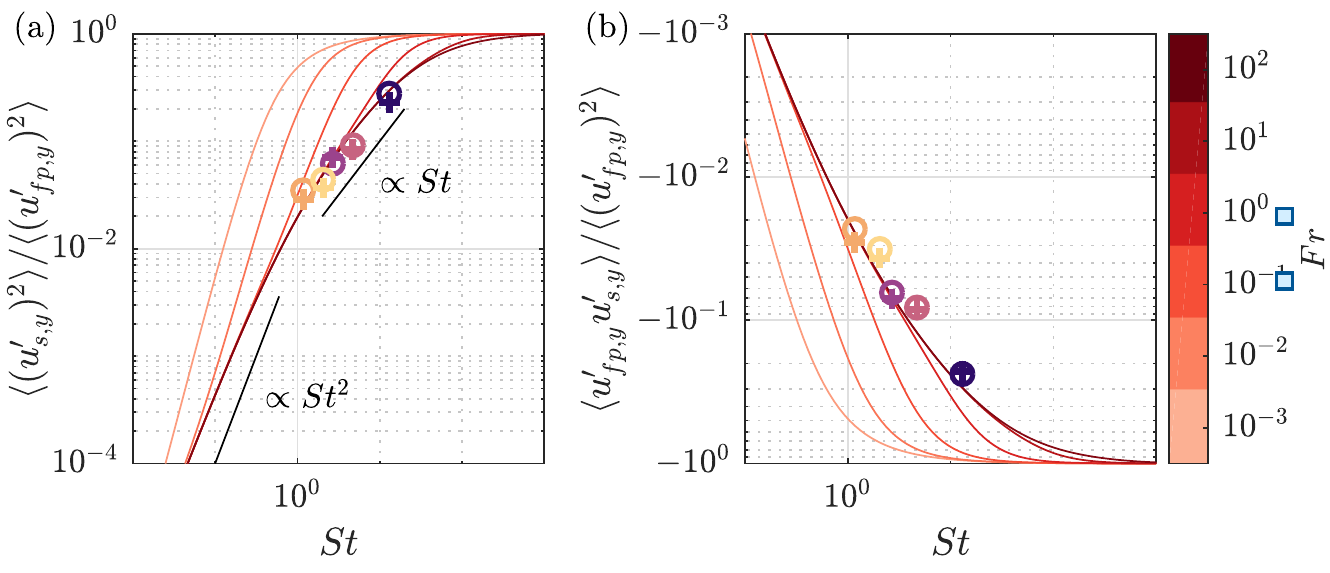}
	\caption{Vertical components of the modeled variance of the slip velocity (a) and covariance of the slip velocity and the sampled fluid velocity (b).
	Red lines indicate the predictions from (\ref{eqn:usmodel}) and $\langle \boldsymbol{u_{fp}}' \boldsymbol{u_{s}}' \rangle = - \langle (\boldsymbol{u_{s}}')^2\rangle$, coloured by $Fr$. Circles indicated measured data, coloured by $Sv$ (see, for example figure~\ref{fig:13} for colour coding). `+' signs indicate case-specific predictions as discussed in the text for matching colours.
	}
	\label{fig:17}
\end{figure}

We now use (\ref{eqn:amodel2}) to derive an expression for the slip velocity variance.
Squaring and averaging the fluctuating part of the particle equation of motion, $\boldsymbol{a_p}' = -\boldsymbol{u_s}'/\tau_p$ , we have $\langle (\boldsymbol{u_s}')^2\rangle = \tau_p^2 \langle (\boldsymbol{a_p}')^2\rangle$.
Substituting in (\ref{eqn:amodel2}) gives:
\begin{equation}
	\frac{\langle(\boldsymbol{u_s}')^2\rangle}{\langle(\boldsymbol{u_{fp}}')^2\rangle} =  \frac{St^2}{(T_{L,fp}/\tau_\eta+St)(T_2/\tau_\eta+St)}
	\label{eqn:usmodel}
\end{equation}
where again we assumed $T_{2,fp} \approx T_2$ and $T_{L,fp}$ is obtained from (\ref{eqn:TLfp}).
Figure~\ref{fig:17}a verifies this prediction, comparing it with the case-specific predictions (using empirical estimates for $T_2$) and the measurements.
The modeled slip velocity variance ranges from $\langle (u_{s,y}')^2 \rangle \approx 0$ in the small $St$ limit, when particles faithfully follow the flow; to $\langle (u_{s,y}')^2 \rangle \approx (u_{fp,y}')^2$ at large $St$, when particles are ballistic and the slip velocity fluctuations effectively equal the velocity fluctuations of the sampled fluid.
In between, the variance of the slip velocity increases with both inertia and gravity as both effects reduce the ability of the particles to follow the fluid fluctuations.
The agreement with the measurements is remarkable.
Also, the generalized model is consistent with the scaling derived by \citet{Balachandar2009} for the different regimes: for the slip velocity variance, his arguments imply $\langle (u_{s,y}')^2 \rangle \propto St^2$ for $\tau_p<\tau_\eta$, $\langle (u_{s,y}')^2 \rangle \propto St$ for $\tau_\eta <\tau_p <T_L$, and $\langle (u_{s,y}')^2 \rangle \approx \mathrm{constant}$ for $\tau_p > T_L$.
Figure~\ref{fig:17}b shows that the covariance of the sampled-fluid velocity and slip velocity is approximately $\langle \boldsymbol{u_{fp}}' \boldsymbol{u_{s}}' \rangle \approx - \langle (\boldsymbol{u_{s}}')^2\rangle$.
This result is supported by the measurements (see figure~\ref{fig:4}), and can be derived from the assumptions of the model: comparing (\ref{eqn:usmodel}) and (\ref{eqn:umodel2}) implies $\langle (\boldsymbol{u_{p}}')^2\rangle = \langle (\boldsymbol{u_{fp}}')^2\rangle - \langle (\boldsymbol{u_{s}}')^2\rangle$, and comparing the latter equality with the variance of the particle velocity (expressed as $\boldsymbol{u_p} = \boldsymbol{u_{fp}} + \boldsymbol{u_s}$) implies $\langle \boldsymbol{u_{fp}}' \boldsymbol{u_{s}}' \rangle = - \langle (\boldsymbol{u_{s}}')^2\rangle$.
Accordingly, the covariance ranges from approximately zero at small $St$, where there is no slip velocity, to $\langle \boldsymbol{u_{fp}}' \boldsymbol{u_{s}}' \rangle = \langle (\boldsymbol{u_{fp}}')^2\rangle$ at large $St$, where the slip velocity equals the sampled fluid velocity.

\begin{figure}
	\centering
	\includegraphics[]{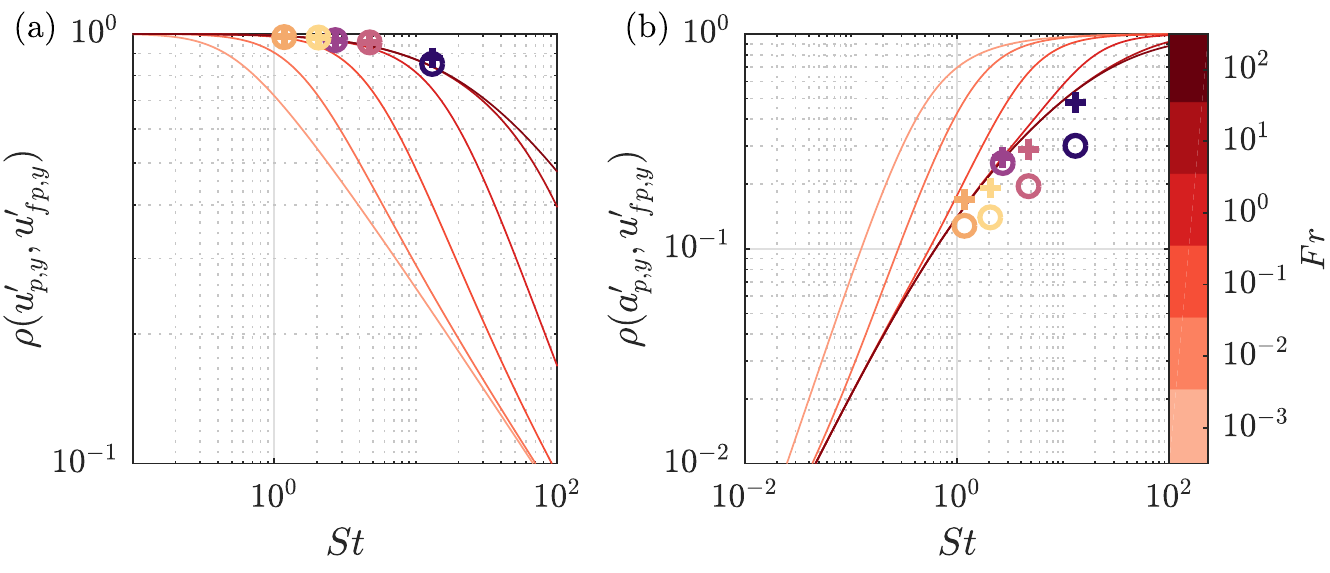}
	\caption{Modeled correlation coefficients of the vertical components of the particle velocity and sampled fluid velocity (a) and of the particle acceleration and sampled fluid velocity (b).
		Red lines indicate the predictions from (\ref{eqn:cov_up_ufp}) and (\ref{eqn:cov_ap_ufp})), coloured by $Fr$. Circles indicated measured data, coloured by $Sv$ (see, for example figure~\ref{fig:13} for colour coding). `+' signs indicate case-specific predictions as discussed in the text for matching colours.
	}
	\label{fig:18}
\end{figure}

From the relationship $\langle \boldsymbol{u_{fp}}' \boldsymbol{u_{s}}' \rangle \approx - \langle (\boldsymbol{u_{s}}')^2\rangle$ (and substituting $\boldsymbol{u_s}'=\tau_p \boldsymbol{a_p}'$ and/or $\boldsymbol{u_s}'=\boldsymbol{u_p}'-\boldsymbol{u_{fp}}'$) we can derive other covariances between the particle velocity, the sampled-fluid velocity, and the particle acceleration; those are especially useful to formulate stochastic models \citep{Zamansky2013,Pope2014}.
Normalizing by the respective rms values, we have expressions for the following correlation coefficients:
\begin{equation}
	\rho(\boldsymbol{u_p u_{fp}}) = \sqrt{1-\frac{St^2}{(T_{L,fp}/\tau_\eta+St)(T_2/\tau_\eta+St)}}
	\label{eqn:cov_up_ufp}
\end{equation}
\begin{equation}
\rho(\boldsymbol{a_p u_{fp}}) = \sqrt{\frac{St^2}{(T_{L,fp}/\tau_\eta+St)(T_2/\tau_\eta+St)}}
\label{eqn:cov_ap_ufp}
\end{equation}
\begin{equation}
\rho(\boldsymbol{a_p u_p}) = 0
\label{eqn:cov_ap_up}
\end{equation}
The model prediction in (\ref{eqn:cov_up_ufp}) and (\ref{eqn:cov_ap_ufp}) are presented in figure~\ref{fig:18} for the vertical component.
The correlation between the particle velocity and sampled-fluid velocity, $\rho(\boldsymbol{u_p u_{fp}})$, approximates unity in the small $St$ limit as expected, and the model predicts the decorrelation due to inertia and gravity (figure~\ref{fig:18}a).
The agreement with the measurements is satisfactory, though a comparison with data at lower $Fr$ is needed to corroborate the prediction.
In contrast, the particle acceleration does not correlate with the fluid velocity at small $St$ (figure~\ref{fig:18}b), with $\rho(\boldsymbol{a_p u_{fp}})$ increasing with gravity and inertia.
Indeed, in the ballistic limit, the slip velocity fluctuations are equal and opposite to the fluid velocity fluctuations, that is $\boldsymbol{a_p}' = -\boldsymbol{u_s}'/\tau_p = \boldsymbol{u_{fp}}'/\tau_p$.
The significant mismatch with the data (about 40~\% for most of the cases) may partly be due to inherent uncertainty on the particle acceleration, but is also likely related to the simplistic assumption that drag and gravity are the only forces at play.
Finally, the prediction that the particle acceleration is uncorrelated from the particle velocity (\ref{eqn:cov_ap_up}) is confirmed by our measurements, from which we find $\rho(\boldsymbol{a_p u_p})<0.05$ for all cases.

\section{Discussion and conclusions} \label{s:conclusion}
We have experimentally investigated the transport of sub-Kolmogorov heavy particles in homogeneous turbulence.
All relevant spatio temporal scales are resolved for the first time in a similar configuration, albeit through planar measurements.
We consider a range of $St$ and $Sv$ for which a rich particle-turbulence interaction is expected, including settling enhancement, inertial filtering, and preferential sampling.
The focus is on the respective roles of inertia and gravity, which have different and often competing influence on the particle motion.
A unique feature of the present measurements is the access to the local properties of the turbulence experienced by the particles along their trajectories.

The importance of the sampled-fluid properties is already clear from the mean vertical velocities.
It is proposed that, in the present range of $St$ and $Sv$, the particle settling velocity can be rationalized by assuming $\langle u_{fp,y} \rangle/u_\eta = C$.
Because, to first order, the settling enhancement $\Delta u_y = \langle u_{fp,y}\rangle$, it is also proportional to the velocity scale of the turbulence, consistent with previously reported trends.
The limited range of $Re_\lambda$, however, cannot clarify whether this scale shall be $u_\eta$, $u_\mathrm{rms}$, or a multi-scale quantity between those.
The mean of the instantaneous slip velocity is well represented by Stokes drag with the Schiller \& Neumann correction, at least for moderate $St$.
Upward- and downward-moving particles sample fluid regions with large velocity fluctuations in those same directions, with similar magnitude in both cases.
Therefore, the settling velocity is augmented due to the downward moving particles being more numerous, not because the sampled downward fluid fluctuations are stronger than the upward ones.

The fluid fluctuations also play a dominant role in determining the particle velocity fluctuations.
The variance of the particle velocity $\langle (\boldsymbol{u_p}')^2 \rangle$ is comparable to but somewhat smaller than the sampled-fluid velocity variance $\langle (\boldsymbol{u_{fp}}')^2 \rangle$, due to inertial filtering; and the latter is slightly smaller than the fluid velocity variance $\langle (\boldsymbol{u_{f}}')^2 \rangle$, due to gravitational drift.
While gravity and inertia have concurrent effects on the particle fluctuating energy, their influences on the particle accelerations are opposite to each other: the crossing-trajectories effect augments the temporal derivative of the sampled-fluid velocity, $\langle \left(\mathrm{d}\boldsymbol{u_{fp}}'/\mathrm{d}t\right)^2 \rangle$, which act to enhance the particle acceleration variance; but this is offset (at least in the present range of parameters) by inertial filtering.
The net result is that heavier particles display smaller rms accelerations and less intermittent acceleration PDFs.
The preferential sampling of high-strain/low-vorticity regions is measurable, but its global impact on the particle motion is weak.

The competing influences of inertia and gravity are on display also in the two-particle statistics.
The uncorrelated component of the relative motion augments the particle velocity structure functions at small separations; while the reduced fluctuating energy of the particles (compared to tracers) has an opposite effect at inertial-range and large-scale separations.
The large relative velocities of nearby particles, which increase with particle inertia, cause heavier particles to separate faster; still, the mean square separation is generally below the expectation for tracers.
This is attributed to gravity causing the particles to experience fluid velocities that decorrelate faster in time, with respect to zero-gravity conditions.
The inertial particles appear to transition out of the ballistic regime at earlier times compared to tracers, similarly to what recently shown for bubbles rising in homogeneous turbulence.

The planar nature of the measurements limits the full characterization of both the particle motion and the turbulent fluid flow.
This is expected to affect especially the two-particle statistics in the form of a biased sampling of the trajectories, although this may not obscure the apparent trends.
The 3D tracking of the particles would overcome this limitation.
Nevertheless, volumetric techniques are presently not capable of simultaneously capturing both phases at the required resolution: despite fast-paced advances in this area, the spatial resolution in volumetric PIV/PTV is still generally below what can be achieved by planar measurements \citep{Discetti2018}; nor these methods have systematically been adapted to multi-phase flows yet.
The present study indeed highlights the central role of the sampled-fluid properties, and extending such an analysis to three-dimensional experiments remains a challenge.

Based on these experimental observations, we have derived an analytical model of particle velocity and acceleration inspired by the seminal work of \citet{Csanady1963}.
This is based on applying a response function to the spectrum of fluid velocities experienced by the particles.
To this end, we use the expression proposed by \citet{Sawford1991} for the fluid velocity autocorrelation, in which we substitute estimates for the time scales $T_L$ and $T_2$ of the sampled fluid.
We lack an analytical expression for the latter time scale, but we show that using the unconditional-fluid formulation (with classic estimates of the $C_0$ constant) has a small quantitative influence.
In its basic form ((\ref{eqn:umodel2}) and (\ref{eqn:amodel2})) the model provides the particle velocity and acceleration variances as a function of $\langle (\boldsymbol{u_{fp}}')^2\rangle$.
The model agrees generally well with the experimental observations, captures the respective effects of inertia and gravity over a wide range of the controlling parameters, and predicts correlations between particle and sampled-fluid velocities and accelerations.
In particular, consistent with the arguments by \citet{Balachandar2009}, it predicts the variance of the slip velocity $\langle (\boldsymbol{u_{s}}')^2\rangle$ to scale as $St^2$ for $\tau_p<\tau_\eta$, as $St$ for $\tau_\eta<\tau_p<T_L$, and to plateau for $\tau_p > T_L$.

Because we show that  $\langle (\boldsymbol{u_{fp}}')^2\rangle$ is only slightly smaller than $\langle (\boldsymbol{u_{f}}')^2\rangle$ (at least in the present range of parameters), the model can be written in a weaker form with more predictive power by substituting $\langle ({u_{fp}}')^2\rangle = u_\eta^2 Re_\lambda/\sqrt{15}$ \citep{Hinze1975}:
\begin{equation}
\frac{\langle(\boldsymbol{u_p}')^2\rangle}{u_\eta^2} = \frac{Re_\lambda}{\sqrt{15}}\left[1- \frac{St^2}{(T_{L,fp}/\tau_\eta+St)(T_2/\tau_\eta+St)}\right]
\end{equation}
\begin{equation}
\frac{\langle(\boldsymbol{a_p}')^2\rangle}{a_\eta^2} = \frac{Re_\lambda}{\sqrt{15}}\frac{St^2}{(T_{L,fp}/\tau_\eta+St)(T_2/\tau_\eta+St)}
\end{equation}
in which $T_2/\tau_\eta$ is a function of $Re_\lambda$ only and $T_{L,fp}/\tau_\eta$ is a function of $Re_\lambda$ and $Sv$ only, given by (\ref{eqn:T2}) and (\ref{eqn:TLfp2}) respectively.
When compared with the present experiments and recent simulations, in particular for the particle acceleration, this version of the model also agrees well with the observations and represents the complex dependency with inertia and gravity.
In particular, it predicts the increase in rms particle acceleration with gravitational drift, and its non-monotonic dependence with $St$ when $Fr \ll 1$ as recently reported by \citet{Ireland2016b}.

Taken together, the laboratory observations and the derived model indicate how, unless the turbulence acceleration is overwhelming ($Fr \gg 1$), both inertia and gravity are key ingredients to understand the transport of heavy particles in homogeneous turbulence.
This calls into question the practice of setting gravity to zero to isolate inertial effects.
Such a consideration appears to be broadly applicable: recent studies on bubbles in homogeneous turbulence \citep{Mathai2016} and heavy particles in turbulent boundary layers \citep{Baker2021} reached a similar conclusion.

\section*{Acknowledgements}
The present work was supported in part by the US Army Research Office, Division of Earth Materials and Processes (grant W911NF-17-1-0366), and Division of Fluid Dynamics (grant W911NF-18-1-0354).

\section*{Declaration of Interests}
The authors report no conflict of interest.

\bibliographystyle{jfm}

\end{document}